%% file: __main_fQ_num.tex
\documentclass[pdftex,twocolumn,epjc3]{svjour3}
\RequirePackage[T1]{fontenc}
\RequirePackage[english]{babel} 
\smartqed     
\usepackage{subcaption}
\usepackage[utf8]{inputenc}
\usepackage{amsmath}      
\usepackage{tabularx}      
\usepackage{booktabs}  
\usepackage{array}
\RequirePackage{graphicx}
\RequirePackage{mathptmx}      
\RequirePackage{balance}       
\RequirePackage{microtype}     

\RequirePackage[colorlinks,citecolor=blue,urlcolor=blue,linkcolor=blue]{hyperref}
\usepackage{float}
\journalname{Eur. Phys. J. C}
\usepackage{adjustbox}
\begin{document}
\sloppy  
\title{Testing $f(Q)$ Gravity with DESI DR2 and Strong-Lensing Time Delays}       

\author{Darshan Kumar$^1$
        \and
        Saibal Ray$^2$
        \and  
        Fengge Zhang$^1$
        \and
        Nisha Rani$^3$
        \and  
        Praveen Kumar Dhankar$^4$
        \and
        Jie Zheng$^5$
}

\institute{Institute for Gravitational Wave Astronomy, Henan Academy of Sciences, Zhengzhou, 450046, China\label{addr1} \email{kumardarshan@hnas.ac.cn; zhangfengge@hnas.ac.cn}
          \and
          Center for Cosmology, Astrophysics and Space Science (CCASS), GLA University, Mathura 281406, Uttar Pradesh, India\label{addr2}
          \and
          Miranda House, University of Delhi, University Enclave, Delhi 110007, India\label{addr3}
          \and
          Symbiosis Institute of Technology, Nagpur Campus, Symbiosis International (Deemed University), Pune 440008, Maharastra, India\label{addr4}
          \and
          School of Microelectronics, Qingdao University of Science and Technology, Qingdao, China\label{addr4}
}
  
\date{Manuscript to be submitted to Eur.Phys.J.C.}    
\maketitle

\newcommand{\btoclnu}{b\to c l\bar{\nu}_{l}}
\newcommand{\btoctaunu}{b\to c\tau\bar{\nu}_{\tau}}
\newcommand{\BtoDorDsttaunu}{\bar{B}\to D^{(*)}\tau\bar{\nu}_{\tau}}
\newcommand{\BtoDorDstellnu}{\bar{B}\to D^{(*)}\ell\bar{\nu}_{\ell}}
\newcommand{\tautomuLHCb}{\tau\to\mu\bar\nu_\mu\nu_\tau}
\newcommand{\tautohadLHCb}{\tau\to 3\pi\nu_\tau}
\newcommand{\tautomuBelle}{\tau\to \ell\bar\nu_{\ell}\nu_{\tau}}
\newcommand{\tautohadBelle}{\tau\to \pi\nu_{\tau}}
\newcommand{\BtoDtaunu}{\bar{B}\to D\tau\bar{\nu}_{\tau}}
\newcommand{\BtoDsttaunu}{\bar{B}\to D^*\tau\bar\nu_{\tau}}
\newcommand{\BtoDststellnu}{\bar{B}\to D^{**}\ell\bar\nu_{\ell}}
\newcommand{\BtoD}{B\!\to\!D}
\newcommand{\BtoDst}{B\!\to\!D^*}
\begin{abstract}
\input{00_abstract}

\end{abstract}

\input{01_introduction}

\input{02_theor_framew}

\input{03_data_method}

\input{04_result}

\input{05_discus_concl}

\section*{Acknowledgments}  
\label{acknowledgements}
\input{acknowledgements}
\section*{Data Availability Statement}
All the observational data used in this work are publicly available. The details of the datasets used in the analysis, along with their sources and references, are given in Section~\ref{sec3}, ``Dataset and Methodology''.


\section*{Code Availability Statement}
Code/software will be made available on reasonable request.


\bibliographystyle{spphys}       
\bibliography{main_ref}   


\balance  
\end{document}

%% file: 00_abstract.tex
Symmetric teleparallel gravity provides an alternative description of gravitation in which non-metricity replaces curvature and torsion. Its extension through $f(Q)$ gravity offers a different geometric description of the late-time expansion of the Universe and its accelerated phase. In this work, we investigate two $f(Q)$ models, a normalized power-law model and a square-root exponential model, and test their ability to describe the late-time expansion history. We constrain the model parameters through Markov chain Monte Carlo analyses using Cosmic Chronometer measurements, DESI DR2 baryon acoustic oscillations, strong-lensing time-delay observations, and three Type Ia supernova compilations, Pantheon$^+$, Union 3.0, and DES Y5. We compare both models with the flat $\Lambda$CDM model using the minimum $\chi^2$, Akaike information criterion, and Bayesian information criterion. The square-root exponential model provides a better statistical fit than $\Lambda$CDM for the combinations of Cosmic Chronometer, DESI DR2, and strong-lensing time-delay data with Pantheon$^+$ and Union 3.0, with improvements in both the goodness of fit and information criteria. The normalized power-law model remains statistically competitive with $\Lambda$CDM for the supernova-inclusive combinations, although the information criteria do not favor its additional parameter. We also determine the transition redshift from cosmic deceleration to acceleration for both models, obtaining consistent values across the different dataset combinations. The transition redshifts agree with observational estimates of the cosmic acceleration epoch. Overall, our results support $f(Q)$ gravity as a viable alternative to $\Lambda$CDM for explaining the late-time accelerated expansion of the Universe without requiring a cosmological constant. 

%% file: 01_introduction.tex
\section{Introduction}\label{sec1}

General Relativity (GR), proposed by Einstein \cite{einstein1916foundation,einstein1922general}, has been remarkably successful in describing gravity and the large-scale evolution of the Universe. However, the discovery of the late-time accelerated expansion of the Universe \cite{1998AJ....116.1009R,1999ApJ...517..565P} has raised fundamental questions about the nature of the mechanism responsible for cosmic acceleration. The evidence for this phenomenon has been supported by several independent observations, including measurements of the Cosmic Microwave Background and the Large Scale Structure of the Universe \cite{2003ApJS..148..233P,2004PhRvD..69j3501T,2005PhRvD..71j3515S,2007ApJS..170..377S,2009ApJS..180..330K,2011ApJS..192...18K}. There are two broad approaches to explain the accelerated expansion. The first introduces a new component in the energy sector of GR, while the second modifies the gravitational theory itself \cite{2003RvMP...75..559P,2006IJMPD..15.1753C,2008ARA&A..46..385F,2010RvMP...82..451S,2011PhR...509..167C,2011PhR...505...59N,2017PhR...692....1N,2019PhR...796....1H}. In the standard cosmological model, the accelerated expansion is attributed to a cosmological constant, leading to the $\Lambda$CDM model. Although $\Lambda$CDM provides an excellent description of a wide range of cosmological observations, it is associated with several unresolved issues, including the coincidence problem, which concerns the comparable present-day energy densities of matter and the cosmological constant, the cosmological constant problem, which arises from the large discrepancy between its theoretical and observed values, and the Hubble tension between early- and late-Universe measurements of the Hubble constant \cite{1999PhRvL..82..896Z,2000astro.ph..5265W,2015PhR...568....1J,2023ARNPS..73..153K}. The physical origin of cosmic acceleration therefore remains an open question and motivates the study of gravitational theories beyond GR.

Among the extensions of GR, $f(R)$ gravity, in which the Einstein-Hilbert action is generalized through an arbitrary function of the Ricci scalar $R$, was introduced by Buchdahl \cite{buchdahl1970non}. Another modification is the $f(R,T)$ theory proposed by Harko et al. \cite{harko2011f}, in which the gravitational Lagrangian depends on both the Ricci scalar $R$ and the trace of the energy-momentum tensor $T$. These theories modify the geometric description of gravity through different extensions of the Einstein-Hilbert action. More generally, the spacetime connection associated with gravitation can be characterized by curvature, torsion, and non-metricity. This leads to three equivalent geometric descriptions of gravity. The first is based on a metric-compatible connection with non-vanishing curvature and vanishing torsion, as in GR. The second uses a curvature-free and metric-compatible connection with non-zero torsion, as in the teleparallel equivalent of GR \cite{aldrovandi2012teleparallel}. The third description uses a connection with vanishing curvature and torsion but non-zero non-metricity. This is known as symmetric teleparallel gravity \cite{aldrovandi2012teleparallel}. In this description, the gravitational interaction is characterized by the non-metricity scalar $Q$ rather than the Ricci scalar or torsion scalar. The symmetric teleparallel equivalent of GR can then be extended by replacing $Q$ with a general function $f(Q)$, giving rise to $f(Q)$ gravity \cite{2022CQGra..39b5013D,2024PhR..1066....1H,2024PDU....4501538N}.

The $f(Q)$ theory has attracted considerable attention as a modified theory of gravity because GR is recovered for the appropriate choice of the function $f(Q)$, while the cosmological dynamics can differ from those of $\Lambda$CDM for more general forms of $f(Q)$. The resulting cosmological equations retain a relatively simple form and allow the effects of non-metricity on the background expansion to be investigated directly. These properties have motivated a number of studies of the cosmological and observational consequences of $f(Q)$ gravity \cite{2019PhRvD.100j4027L,2020PhRvD.102l4029M,2022EPJC...82..303Z,2023MNRAS.522..252S,bhardwaj2024cosmological,dubey2025study,jasim2025study}. In recent years, several aspects of $f(Q)$ cosmology have been investigated using different theoretical models and observational datasets \cite{2019EPJC...79..530L,2019PhRvD.100j4027L,2020PhRvD.101j3507J,2021PhRvD.103f3505A,2021PhLB..82236634A,2022PhRvD.105h4061E,2023PhRvD.107d4022K,2023EPJC...83...58A,2024EPJC...84..806E,2024MNRAS.533.2232Y,2025arXiv250217937N,2025JCAP...12..011B}. These studies motivate further tests of specific $f(Q)$ forms against the increasingly precise late-time cosmological observations.   

Motivated by the possibility that $f(Q)$ gravity can provide an alternative description of the accelerated expansion, we consider two different forms of $f(Q)$ gravity in this work. The first is a normalized power-law model \cite{2020PhRvD.101j3507J,2019EPJC...79..606D,2021PhRvD.103f3505A,2023PhRvD.107d4022K} given by $f(Q)=Q+\alpha Q_0\left(Q/Q_0\right)^n$, where $n$ controls the deviation from the linear dependence on $Q$. The second is a square-root exponential model \cite{2021PhLB..82236634A,2024ChJPh..89.1754G} given by $f(Q)=Q+\alpha Q_0\left[1-\exp\left(-m\sqrt{Q/Q_0}\right)\right]$, where $m$ characterizes the exponential correction. These two models provide different modifications to the background expansion and allow us to examine whether departures from the standard $\Lambda$CDM evolution are supported by current observations. We constrain their free parameters through Markov chain Monte Carlo analyses using Cosmic Chronometer measurements of the Hubble parameter, baryon acoustic oscillation measurements from DESI DR2, strong gravitational lensing time-delay observations, and Type Ia Supernovae data from the Pantheon$^+$, Union3.0, and DES Y5 compilations. We compare the two $f(Q)$ models with the flat $\Lambda$CDM model using the minimum $\chi^2$, Akaike information criterion, and Bayesian information criterion. We also determine the transition redshift from cosmic deceleration to acceleration and examine its consistency across the different observational combinations.

This paper is structured as follows. Section \ref{sec2} outlines the overview of the mathematical formalism of $f(Q)$ gravity within a spatially flat FLRW background. In Section \ref{sec3}, we describe the observational datasets and the statistical methodology used for model comparison. We present our primary constraints and numerical results in Section \ref{sec4}. Finally, Section \ref{sec5} summarizes the conclusion drawn from the study with a general discussion of our results.

%% file: 02_theor_framew.tex
\section{Theoretical framework}\label{sec2}   
We consider a four-dimensional spacetime described by a metric tensor $g_{\mu\nu}$ and a general affine connection $\Gamma^\alpha_{\mu\nu}$. In symmetric teleparallel gravity, the curvature and torsion associated with the affine connection vanish, while the non-metricity remains non-zero. Thus, the geometric conditions are given by
\begin{equation}
R^{\alpha}{}_{\beta\mu\nu}=0\,, \qquad T^{\alpha}{}_{\mu\nu}=0\,.
\end{equation}
A general affine connection can be decomposed into three independent parts as      
\begin{equation}
\Gamma^\alpha_{\mu\nu}=\tilde{\Gamma}^\alpha_{\mu\nu}+K^\alpha{}_{\mu\nu}+L^\alpha{}_{\mu\nu}\,,
\end{equation}
where $\tilde{\Gamma}^\alpha_{\mu\nu}$ is the Levi--Civita connection, $K^\alpha{}_{\mu\nu}$ is the contortion tensor, and $L^\alpha{}_{\mu\nu}$ is the deformation tensor. In this description, the departure from metric compatibility is characterized by the non-metricity tensor, defined as
\begin{equation}
Q_{\alpha\mu\nu}=\nabla_{\alpha}g_{\mu\nu}=\partial_{\alpha}g_{\mu\nu}-\Gamma^{\sigma}_{\alpha\mu}g_{\sigma\nu}-\Gamma^{\sigma}_{\alpha\nu}g_{\mu\sigma}\,.
\end{equation}

Since both curvature and torsion vanish, a suitable coordinate transformation can always be chosen such that the affine connection vanishes. This choice is known as the coincident gauge \cite{BeltranJimenez:2017tkd}. In this gauge, the covariant derivative reduces to the partial derivative, which simplifies the calculation of the non-metricity tensor.

The gravitational dynamics in symmetric teleparallel gravity are described by the non-metricity scalar $Q$, defined as \cite{1999ChJPh..37..113N}
\begin{equation}
Q=-Q_{\alpha\mu\nu}P^{\alpha\mu\nu}\,,
\end{equation}
where $P^{\alpha}{}_{\mu\nu}$ is the non-metricity conjugate given by
\begin{equation}
P^{\alpha}{}_{\mu\nu}=-\frac{1}{4}Q^{\alpha}{}_{\mu\nu}+\frac{1}{2}Q_{(\mu}{}^{\alpha}{}_{\nu)}+\frac{1}{4}(Q^{\alpha}-\tilde{Q}^{\alpha})g_{\mu\nu}-\frac{1}{4}\delta^{\alpha}_{(\mu}Q_{\nu)}\,.
\end{equation}

The two traces of the non-metricity tensor are defined as
\begin{equation}
Q_{\mu}=Q_{\mu\nu}{}^{\nu}\,, \qquad \tilde{Q}_{\mu}=Q^{\nu}{}_{\mu\nu}\,.
\end{equation}

\subsection{$f(Q)$ Cosmology}\label{Sec3}

The symmetric teleparallel equivalent of GR can be extended by considering an arbitrary function of the non-metricity scalar $Q$. The action of $f(Q)$ gravity is given by \cite{BeltranJimenez:2017tkd}
\begin{equation}\label{equ_action}
S=\int\left[-\frac{1}{2\kappa}f(Q)+\mathcal{L}_{m}\right]\sqrt{-g}\,d^{4}x\,,
\end{equation}
where $f(Q)$ is a function of the non-metricity scalar, $g$ is the determinant of the metric tensor, $\mathcal{L}_{m}$ is the matter Lagrangian, and $\kappa=8\pi G_{\rm N}=1$, with $G_{\rm N}$ denoting the gravitational constant. Varying the action with respect to the metric tensor gives the gravitational field equations
\begin{equation}
\begin{split}
\frac{2}{\sqrt{-g}}\nabla_{\alpha} &
\left(
\sqrt{-g} f_{Q} P^{\alpha}{}_{\mu\nu}
\right) 
+\frac{1}{2} g_{\mu\nu} f \\
& + f_{Q}
\left(
P_{\mu\alpha\beta} Q_{\nu}{}^{\alpha\beta}
-2 Q_{\alpha\beta\mu} P^{\alpha\beta}{}_{\nu}
\right)
= T_{\mu\nu} \, ,
\end{split}
\end{equation}
where $f_Q=df/dQ$ and $T_{\mu\nu}$ denotes the energy-momentum tensor of matter. For a perfect fluid, the energy-momentum tensor is given by
\begin{equation}
T_{\mu\nu}=(\rho+p)u_{\mu}u_{\nu}+pg_{\mu\nu}, \qquad u_{\mu}u^{\mu}=-1\,.
\end{equation}

We consider a homogeneous and isotropic spatially flat FLRW spacetime described by the line element
\begin{equation}
ds^{2}=-dt^{2}+a^{2}(t)\left[dr^{2}+r^{2}(d\theta^{2}+\sin^{2}\theta\,d\phi^{2})\right]\,,
\end{equation}
where $a(t)$ is the scale factor and $H=\dot{a}/a$ is the Hubble parameter. For this metric, the non-metricity scalar reduces to
\begin{equation}
Q=6H^{2}\,.
\end{equation}

The modified Friedmann equations then take the form \cite{2020PhRvD.101j3507J}
\begin{equation}\label{modif_F_eq_1}
3H^{2}=\frac{1}{2f_Q}\left(\rho+\frac{f}{2}\right)\,,
\end{equation}
\begin{equation}
\dot{H}+\left(3H+\frac{\dot{f}_Q}{f_Q}\right)H=\frac{1}{2f_Q}\left(-p+\frac{f}{2}\right)\,.
\end{equation}



For a Universe containing matter and radiation, the total energy density is
\begin{equation}
\rho=3H_0^2\left[\Omega_{m0}(1+z)^3+\Omega_{r0}(1+z)^4\right]\,,
\end{equation}
where $H_0$ is the present-day Hubble constant, expressed in units of $\mathrm{km\,s^{-1}\,Mpc^{-1}}$, and $\Omega_{m0}$ and $\Omega_{r0}$ denote the present-day matter and radiation density parameters. Using $Q=6H^2$ and $Q_0=6H_0^2$, the background equations for the two models can be obtained from Eq.~(\ref{modif_F_eq_1}).

\subsubsection{Model I: Normalized Power-law $f(Q)$}

The first model considered in this work is a normalized power-law form of $f(Q)$,
\begin{equation}
{f_I(Q)=Q+\alpha Q_0\left(\frac{Q}{Q_0}\right)^n}
\end{equation}
where $\alpha$ and $n$ are dimensionless parameters. For this model, the derivative of $f(Q)$ with respect to $Q$ is
\begin{equation}
f_Q=1+\alpha n\left(\frac{Q}{Q_0}\right)^{n-1}\,.
\end{equation}


Substituting this form into Eq.~(\ref{modif_F_eq_1}) gives the background equation
\begin{equation}
{H^2+\alpha(2n-1)H_0^2\left(\frac{H^2}{H_0^2}\right)^n=H_0^2\left[\Omega_{m0}(1+z)^3+\Omega_{r0}(1+z)^4\right]}.
\end{equation}

The present-day condition $H(z=0)=H_0$ fixes $\alpha$ as
\begin{equation}
{\alpha=\frac{\Omega_{m0}+\Omega_{r0}-1}{2n-1}}.
\end{equation}

Thus, after fixing $\Omega_{r0}$ from Planck's 2018 survey \cite{1807.06209}, the background evolution of Model I is determined by the parameters $H_0$, $\Omega_{m0}$, and $n$.

\subsubsection{Model II: Square-root Exponential $f(Q)$}       

The second model considered in this work has a square-root exponential form,
\begin{equation}
{f_{II}(Q)=Q+\alpha Q_0\left[1-\exp\left(-m\sqrt{\frac{Q}{Q_0}}\right)\right]}
\end{equation}
where $\alpha$ and $m$ are dimensionless parameters. For this model, the derivative of $f(Q)$ is
\begin{equation}
f_Q=1+\frac{\alpha m}{2}\sqrt{\frac{Q_0}{Q}}\exp\left(-m\sqrt{\frac{Q}{Q_0}}\right)\,.
\end{equation}

Using $Q=6H^2$ and $Q_0=6H_0^2$, the background equation becomes
\begin{equation}
\begin{split}
H^2+\alpha mHH_0\exp\left(-\frac{mH}{H_0}\right)-&\alpha H_0^2 \left[1-\exp\left(-\frac{mH}{H_0}\right)\right] \\&= H_0^2\left[\Omega_{m0}(1+z)^3 +\Omega_{r0}(1+z)^4\right].
\end{split}
\end{equation}

The present-day condition $H(z=0)=H_0$ gives
\begin{equation}
{\alpha=\frac{\Omega_{m0}+\Omega_{r0}-1}{(m+1)e^{-m}-1}}.
\end{equation}

Hence, after fixing $\Omega_{r0}$, the independent parameters of Model II are $H_0$, $\Omega_{m0}$, and $m$.

For both models, the corresponding background equations are implicit equations for the Hubble parameter $H(z)$. We therefore solve these equations numerically at each redshift for the given values of the model parameters. The resulting $H(z)$ is then used to calculate the cosmological observables required for the subsequent parameter estimation and comparison with the observational datasets.

%% file: 03_data_method.tex
\section{Datasets and Methodology}\label{sec3}

We use recent and complementary observational datasets, that includes Hubble parameter measurements from Cosmic Chronometers, Type Ia Supernovae from Pantheon Plus,  DES Y5 and Union 3.0 compilation, Baryon Acoustic Oscillations from DESI DR2, and Strong Gravitational Lensing Time-Delay Observations in this analyses. Further, we adopt a Bayesian statistics to investigate a power-law form of the $f(Q)$ theory given by $f (Q) = Q + \alpha Q^n$ and explore its cosmological implications in the presence of dynamical dark energy.

\subsection{Hubble parameter (Cosmic Chronometers)}

The Hubble parameter derived from Cosmic Chronometers (CC) provides a direct and model-independent measurement to probe the cosmic expansion history, based on the differential ages of passively evolving galaxies. The method was first proposed by Jimenez and Loeb \cite{2003ApJ...593..622J,2005PhRvD..71l3001S}. In this method, the Hubble parameter in terms of cosmological redshift can be written as
\begin{equation}\label{hubble_parameter}
    H(z)=-\frac{1}{(1+z)}\frac{dz}{dt}.
\end{equation}

In Eq. \eqref{hubble_parameter}, the term $dt$ is estimated using the differential age evolution of the
universe in a given interval of redshift $dz$. For astrophysical objects, one can measure the redshift directly with uncertainty close to zero.
The other factor that needs to be measured is the differential age of galaxies, $dt$. This differential age measurement can be done using various methods like absorption feature analyses, full spectrum fitting and calibration of specific spectroscopic features \cite{1994ApJS...95..107W,2011MNRAS.412.2183T,2012JCAP...08..006M}. But it is important to take precautions while selecting the galaxies for this purpose. It is because in the young evolving galaxies, stars are continually born so  their emission spectra is dominated by young stellar population. Hence for the accurate estimation of the $dt$, passively evolving red galaxies whose spectra is mainly dominated by the old stellar population need to be selected. In this work, we use Hubble parameter measurements having 32 data points as well as its covariance matrix given in \cite{2022LRR....25....6M}. Throughout this work, this dataset is denoted as \emph{\bf CC}.

\subsection{Type Ia supernovae}

Type Ia Supernova (SNIa) data represent one of the most powerful observational probes of the cosmic expansion history. Their remarkably uniform peak luminosities allow standardization through light-curve shape and color corrections, making SNIa reliable standard candles. SNIa observations provide measurements of the luminosity distance as a function of redshift through the distance modulus-redshift relation. The distance modulus, $\mu_{SNe}$, defined as the difference between the apparent and absolute magnitudes of a supernova, is expressed as
\begin{equation}\label{distance modulus}
    \mu_{SNe}(z)=m_B(z)+\alpha . X_1- \beta . C-M_B,
\end{equation}
where $m_B$ is peak magnitude in the rest frame B-band and $M_B$ represents absolute magnitude measured through B-band 
of a fiducial Type Ia SNe with $X_1 = 0$ and $C = 0$. Here $C$ and $X_1$ represent the supernova color at maximum brightness and time stretch of light curve respectively. 

A large, homogeneous compilations such as Union, JLA, Pantheon, Pantheon Plus, and Union 3.0 span a wide redshift range from the local Universe to $z \geq 2$, with controlled systematic uncertainties. Out of these, we used the three different compilation of observed SNIa namely Pantheon Plus, Union 3.0 and DES Y5. These datasets are collectively referred to as \emph{\bf SNIa}.

\subsubsection{Pantheon Plus}

The Pantheon Plus is compilation of 18 different samples, where a sample means the data produced by a single Supernova survey over a discrete interval of time \cite{2022ApJ...938..113S}. It includes observations from multiple major surveys like Pan-STARRS1 (PS1), SDSS, SNLS, HST, and several well-calibrated low-redshift samples. One of the key advancement of Pantheon Plus is uniform reanalyses of light curves with improved photometric calibration, forward modeling of the supernova and a comprehensive treatment of systematic uncertainties \cite{2021ApJ...909...26B,2021ApJ...913...49P,2023ApJ...945...84P}. It contains 1701 spectroscopically confirmed SNIa spanning a wide range of redshift, $0.001\leq z \leq 2.261$. Public release of the Pantheon Plus provides likelihood function with the systematic and statistical covariance matrices \cite{2019ApJ...874..150B}. In order to reduce the peculiar velocity effect on the Hubble diagram, data with $z\leq 0.01$ is commonly not considered in the cosmological analyses \cite{2022ApJ...938..112P}. Keeping this in mind, we exclude data points below 0.01 redshift. Therefore, we work with 1590 data points out of available 1701 points. In the following, the Pantheon Plus Supernova sample is referred to as \emph{\bf Pantheon$^+$}.

\subsubsection{Union 3.0}

The Union 3.0 compilation comprises an extensive sample of 2087 Type Ia supernovae \cite{2025ApJ...986..231R}. Out of these 2087 data points, 1363 are common with the Pantheon Plus dataset. Unlike Pantheon Plus, Union 3.0 employs a distinct statistical framework for handling systematic uncertainties, based on Bayesian hierarchical modeling.  This allows for a coherent treatment of measurement errors and observational biases, making Union 3.0 especially valuable for cross-checking results obtained from Pantheon Plus and for performing independent cosmological analyses. For our purpose, we use the most recent Union 3.0 release, covering the redshift interval  $0.01<z<2.26$. Our analyses is confined to the binned distance modulus data which are publicly available for this compilation at present. We also consider their corresponding covariance matrix to ensure a consistent treatment of uncertainties.

\subsubsection{DES Y5}

 The DES Y5 supernova dataset is taken from the complete five-year observational campaign of the Dark Energy Survey. It is a uniformly selected sample optimized for precision cosmology and consists of 1635 Type Ia supernovae identified through photometric classification across the redshift interval $0.0596<z<1.12$, along with 194 SNIa at $z<0.1$ obtained from recent low-redshift surveys \cite{2024ApJ...973L..14D}. In order to reduce calibration-related systematics, DES Y5 excludes older legacy low-redshift samples that are used in other compilations. As for most of the DES SNe spectroscopic confirmation is unavailable, so a probabilistic classification using the Supernova Bayesian framework is adopted \cite{Kessler_2017,2020MNRAS.491.4277M}. The resulting uncertainties in SNe type assignment are incorporated into the covariance matrix used for cosmological analyses. Compared to Pantheon Plus, the DES Y5 sample is believed to provide enhanced constraints in the dark energy dominated epoch as it contains a substantially higher proportion of supernovae at  $z>0.5$. The combination of a homogeneous survey strategy, modern calibration, and strong high-redshift coverage makes DES Y5 a powerful and complementary dataset for investigating the cosmological models.
 
\subsection{Baryon Acoustic Oscillations (DESI DR2)} 

Baryon Acoustic Oscillations (BAO) constitute one of the most established statistical standard rulers used in observational cosmology \cite{2010dken.book..246B}. Their origin traces back to acoustic perturbations in the tightly coupled photon--baryon fluid of the early universe. During this phase, radiation pressure opposed gravitational collapse until the epoch of recombination, at which point photons decoupled and formed the Cosmic Microwave Background (CMB). The baryonic component preserved a residual signature of these acoustic modes, which appears as a preferred clustering scale in the late-time matter distribution. This characteristic scale corresponds to the comoving sound horizon \cite{1998ApJ...496..605E,1972CoASP...4..173S}. The same physical scale is observed across several independent tracers, including the galaxy two-point correlation function, the galaxy power spectrum, and the CMB anisotropy spectrum \cite{1970ApJ...162..815P}. The large-scale nature and well-understood physical origin of this feature make it a reliable probe of the cosmic expansion history \cite{2007ApJ...664..675E,2012MNRAS.426.1280S}.        

In this work, we use the second-year BAO measurements released by the Dark Energy Spectroscopic Instrument (DESI DR2), which represent three years of accumulated observations \cite{2025PhRvD.112h3511L,2025PhRvD.112h3515A}. The reported measurements include the full covariance matrix, which captures correlations among the various distance indicators and allows for a consistent statistical treatment. As a result, the DESI BAO dataset provides precise distance constraints that are well suited for cosmological parameter inference. 

The BAO observables are expressed in terms of three distance ratios, each normalized by the sound horizon at the drag epoch, $r_d$. These ratios correspond to the transverse comoving distance $d_\mathrm{co}(z)/r_d$, the Hubble distance $d_H(z)/r_d$, and the volume-averaged distance $d_V(z)/r_d$, which are defined as
\begin{align}
    \text{First quantity}: \quad & \dfrac{d_\mathrm{co}(z)}{r_d} = \dfrac{d_L(z)}{r_d\,(1+z)}, \label{eq_dM} \\
    \text{Second quantity}: \quad & \dfrac{d_H(z)}{r_d} = \dfrac{c}{r_d\,H(z)}, \label{eq_dH} \\
    \text{Third quantity}: \quad & \dfrac{d_V(z)}{r_d} = \dfrac{\left[z\,d_H(z)\,d_\mathrm{co}^2(z)\right]^{1/3}}{r_d}. \label{eq_dV}
\end{align}

The sound horizon $r_d$ depends on the physical matter and baryon densities, as well as on the effective number of relativistic species. Through this dependence, BAO distance measurements connect late-time structure to early-universe physics. 

BAO observables show a strong degeneracy between the Hubble constant $H_0$ and the sound horizon $r_d$, which prevents a simultaneous determination of both parameters from BAO data alone. To address this limitation, we adopt a prior on the baryon density parameter, $\Omega_{b0} = 0.02218 \pm 0.00055$, which is used in the calculation of $r_d$ in our analyses \cite{2025JCAP...02..021A}. This choice breaks the degeneracy and permits robust cosmological constraints. Under this assumption, the DESI DR2 BAO measurements provide an independent distance probe that complements other cosmological datasets. In this analyses, we impose a Gaussian prior on the sound horizon $r_d$. This compilation is denoted as \emph{\bf DESI DR2}

\subsection{Strong Gravitational Lensing (Time Delay)}

In strong gravitational lensing systems, time delays between multiple images provide a direct, model-independent method to constrain cosmological parameters. Light emitted simultaneously from a background source reaches the observer at different times because each ray follows a distinct path and passes through a different gravitational potential, which produces measurable time delays between the images. When the source varies intrinsically, these delays can be determined by monitoring the brightness of the lensed images, as each variation corresponds to the same underlying event. Within this framework, the time-delay distance plays a central role in the determination of cosmological parameters. This distance depends on a combination of three angular diameter distances: the observer–lens distance $d_\mathrm{A}^{^\mathrm{ol}}$, the lens–source distance $d_\mathrm{A}^{^\mathrm{ls}}$, and the observer–source distance $d_\mathrm{A}^{^\mathrm{os}}$.

For a given source position $\mathcal{B}$ and image position $\theta$, the time-delay difference $\Delta t$ between two images, labeled $i$ and $j$, is expressed as \cite{1990CQGra...7.1319P,1990CQGra...7.1849P}:
\begin{equation}\label{equ_time_diff}
\begin{split}
\Delta t_{ij} \equiv \delta t_j - \delta t_i = \frac{(1+z_l)}{c} 
\frac{d_\mathrm{A}^{^\mathrm{os}} d_\mathrm{A}^{^\mathrm{ol}}}{d_\mathrm{A}^{^\mathrm{ls}}} 
\Bigg[ & \frac{(\theta_j - \mathcal{B})^2}{2} - \psi(\theta_j) \\
& - \frac{(\theta_i - \mathcal{B})^2}{2} + \psi(\theta_i) \Bigg],
\end{split}
\end{equation}
where $z_l$ denotes the lens redshift and $\psi$ represents the lensing potential of the foreground galaxy. 
  
From Eq~\ref{equ_time_diff}, the time-delay distance, $d_{\Delta t}$, can be expressed as
\begin{equation}\label{equ_tdd}
d_{\Delta t} = (1+z_l)\frac{d_\mathrm{A}^{^\mathrm{os}} d_\mathrm{A}^{^\mathrm{ol}}}{d_\mathrm{A}^{^\mathrm{ls}}}.
\end{equation}

The determination of the time delay distance requires three key components. First, precise measurements of the time delays between lensed images. Second, a careful account of lensing contributions from mass structures along the line of sight. Third, a reliable mass model for the lens galaxy itself. The lens galaxy potential governs the observed time delays, but additional matter along the line of sight can focus or de-focus light rays, producing systematic shifts in the inferred time-delay distance \cite{1994ApJ...436..509S,2016A&ARv..24...11T}. This effect is typically quantified through the external convergence, $\kappa_{\mathrm{ext}}$, which modifies both $d_{\Delta t}$ and the Hubble constant $H_0$. Although the mean value of $\kappa_{\mathrm{ext}}$ vanishes across the sky, neglecting this contribution in individual systems can bias measurements of $d_{\Delta t}$ and $H_0$. In addition, lens modeling alone cannot fully constrain $\kappa_{\mathrm{ext}}$ due to the mass-sheet degeneracy. The analyses of line-of-sight structures in an independent manner is therefore necessary to obtain robust constraints \cite{2003ApJ...584..664K,2014MNRAS.443.3631M}.

In this work, we consider the H0LiCOW ($H_0$ Lenses in COSMOGRAIL's Wellspring) sample, which includes seven lens systems: B1608+656 \cite{2010ApJ...711..201S,2019Sci...365.1134J}, RXJ1131-1231 \cite{2013ApJ...766...70S,2014ApJ...788L..35S,2019MNRAS.490.1743C}, HE 0435-1223 \cite{2017MNRAS.465.4895W,2019MNRAS.490.1743C}, SDSS 1206+4332 \cite{2019MNRAS.484.4726B}, WFI2033-4723 \cite{2020MNRAS.498.1440R}, PG 1115+080 \cite{2019MNRAS.490.1743C}, and DES J0408-5354 \cite{2017ApJ...838L..15L,2020MNRAS.494.6072S}. These systems provide high-quality measurements of time delays and lens models, allowing precise determinations of the time-delay distances and constraints on the Hubble constant.  The time delay measurements are labeled as \emph{Time Delay} throughout this paper.

\subsection{Statistical Analyses}

In this work, we carry out a Bayesian parameter inference by comparing theoretical predictions with observational measurements. The analyses proceeds through the construction of chi-square functions for each dataset, followed by their combination into a joint likelihood. The datasets included in this study consist of Hubble parameter measurements from Cosmic Chronometers, Type Ia Supernovae (SNIa), Baryon Acoustic Oscillations (BAO), and time-delay distance measurements from Strong Gravitational Lensing systems (SGL). The statistical treatment adopted for each dataset is described below. 

\subsubsection{Hubble parameter (Cosmic Chronometer)} 

For the Cosmic Chronometer dataset, we define the chi-square statistic as
\begin{equation}\label{eq_chi_1}
\begin{split}
\chi^{2}_{\mathrm{CC}} = &
\left[
\mathrm{H}_{\mathrm{th}}
- \mathrm{H}_{\mathrm{obs}}\left(z_i\right)
\right]^T  \operatorname{Cov}_{ij}^{-1}
\left[
\mathrm{H}_{\mathrm{th}}
- \mathrm{H}_{\mathrm{obs}}\left(z_j\right)
\right].
\end{split}
\end{equation}

Here, $\mathrm{H}_{\mathrm{th}}$ and $\mathrm{H}_{\mathrm{obs}}$ denote the theoretical and observed values of the Hubble parameter, respectively. The theoretical prediction $\mathrm{H}_{\mathrm{th}}$ follows from the modified gravity scenarios considered in this work. The matrix $\operatorname{Cov}_{ij}$ represents the total uncertainty associated with the Hubble parameter measurements and incorporates both statistical and systematic contributions. In our analyses, we account for the full covariance structure through
\begin{equation}\label{cov_1}
\operatorname{Cov}_{ij}
=
\operatorname{Cov}_{ij}^{\mathrm{stat}}
+
\operatorname{Cov}_{ij}^{\mathrm{syst}},
\end{equation}
where $\operatorname{Cov}_{ij}^{\mathrm{stat}}$ denotes the contribution from statistical uncertainties, while $\operatorname{Cov}_{ij}^{\mathrm{syst}}$ accounts for systematic effects. The systematic covariance follows the detailed assessment presented by ~\cite{2022LRR....25....6M}. In particular, four dominant sources of systematic uncertainty are included in the covariance construction, namely the initial mass function, the stellar library, the treatment of metallicity, and the choice of stellar population synthesis models \cite{2020ApJ...898...82M}. A comprehensive discussion of the full covariance matrix and its construction can be found in Section~3.1.4 of ~\cite{2022LRR....25....6M}.  \\

\subsubsection{Strong Gravitational Lensing (Time Delay)}

The chi-square associated with the time-delay distance measurements from  H0LiCOW lenses is defined as
\begin{equation}\label{equ_chi_sgl}
    \chi^2_{\mathrm{SGL}}
    = \sum_i
    \left[
    \frac{
    d_{\Delta t}^{\mathrm{th}}
    - d_{\Delta t}^{\mathrm{obs}}
    }{
    \sigma_{d_{\Delta t}^{\mathrm{obs}}}
    }
    \right]^2 ,
\end{equation}
where $d_{\Delta t}^{\mathrm{th}}$ denotes the theoretical prediction for the time-delay distance, as defined in equation~\eqref{equ_tdd}. The observed time-delay distance and its corresponding uncertainty are denoted by $d_{\Delta t}^{\mathrm{obs}}$ and $\sigma_{d_{\Delta t}^{\mathrm{obs}}}$, respectively.

\subsubsection{Type Ia Supernovae} 

The theoretical apparent magnitude for Type Ia supernovae is written as
\begin{align}\label{equ_m_th}
    \mu^{\mathrm{th}}
    &= 5\log_{10}\left(\dfrac{d_L}{1\,\mathrm{Mpc}}\right) + 25  
\end{align}        

The chi-square for the SNIa dataset takes the form  
\begin{equation}
    \chi^2_{\mathrm{SNIa}}=\Delta\bar{\mu}^T \cdot \mathrm{Cov}^{-1} \cdot \Delta\bar{\mu}
\end{equation}
with
\begin{align}
    \Delta\bar{\mu} &= \mu^{\mathrm{obs}} - \mu^{\mathrm{th}},
\end{align}
where $\mathrm{Cov}$ denotes the full covariance matrix.

\subsubsection{Baryon Acoustic Oscillations}

The chi-square function associated with the BAO dataset is written as
\begin{equation}
\begin{split}
    \chi^2_{\mathrm{BAO}}
    = &
    \left[
    X_{\mathrm{th}}
    - X_{\mathrm{obs}}\left(z\right)
    \right]^T
    \mathrm{Cov}^{-1} 
    \left[
    X_{\mathrm{th}}
    - X_{\mathrm{obs}}\left(z\right)
    \right],
\end{split}
\end{equation}
where the vector $X$ contains the BAO observables defined in Equations~(\ref{eq_dM}), (\ref{eq_dH}), and (\ref{eq_dV}).\\

\subsubsection{Total Chi-square and Likelihood}

The total chi-square combines contributions from all datasets as
\begin{equation}
    \chi^2_T
    = \chi^{2}_{\mathrm{CC}}
    + \chi^2_{\mathrm{SGL}}
    + \chi^2_{\mathrm{SNIa}}
    + \chi^2_{\mathrm{BAO}}.
\end{equation}

The supernova contribution corresponds to the specific samples used in this work, namely $\chi^2_{\mathrm{PantheonPlus}}$, $\chi^2_{\mathrm{Union3}}$, and $\chi^2_{\mathrm{DES Y5}}$. 

The likelihood function is defined as
\begin{equation}
    \mathcal{L} \propto \exp\left(-\frac{\chi^2_T}{2}\right).    
\end{equation}

Since the datasets are statistically independent, the individual chi-square terms add directly. This approach allows a joint inference of cosmological parameters within a unified statistical framework. 

We note that the radiation density parameter, $\Omega_{r0}$, is omitted due to its negligible contribution relative to the dominant components, and this omission does not alter the MCMC results or the conclusions of this work.

%% file: 04_result.tex
\section{Results}\label{sec4}

We present the constraints on the two $f(Q)$ gravity models using the combinations of Cosmic Chronometer (CC), DESI DR2, strong gravitational lensing time-delay, and Type Ia Supernovae data. The supernova compilations considered here are Pantheon$^+$, Union 3.0, and DES Y5. We first discuss the constraints obtained for the normalized power-law model $f_{\rm I}(Q)$, followed by those for the square-root exponential model $f_{\rm II}(Q)$. We then compare both models with the flat $\Lambda$CDM model using the minimum chi-square, Akaike Information Criterion (AIC), and Bayesian Information Criterion (BIC). Finally, we determine the transition redshift from cosmic deceleration to acceleration for both models.

\subsection{Constraints on the normalized power-law model}\label{sec_results_fQI}

Table~\ref{tab:mcmc_constraints_fQI} presents the marginalized constraints on $H_0$, $\Omega_{m0}$, and $n$ for the normalized power-law model $f_{\rm I}(Q)$ at the $68\%$ confidence level. For the combination of CC + DESI DR2 + Time-delay, we obtain $H_0=68.9139^{+0.6063}_{-0.6187}$, $\Omega_{m0}=0.2916^{+0.0067}_{-0.0069}$, and $n=0.0460^{+0.1023}_{-0.1096}$. The inclusion of the supernova compilations leads to relatively tighter constraints on $H_0$ and $\Omega_{m0}$. The value of $n$ also shifts towards positive values when the Type Ia Supernovae data are included, with the largest value obtained for the Union 3.0 combination. The constraints from Pantheon$^+$ and DES Y5 remain close to each other, while Union 3.0 prefers a somewhat larger value of $n$.

\begin{figure}
    \centering
    \includegraphics[width=1\linewidth]{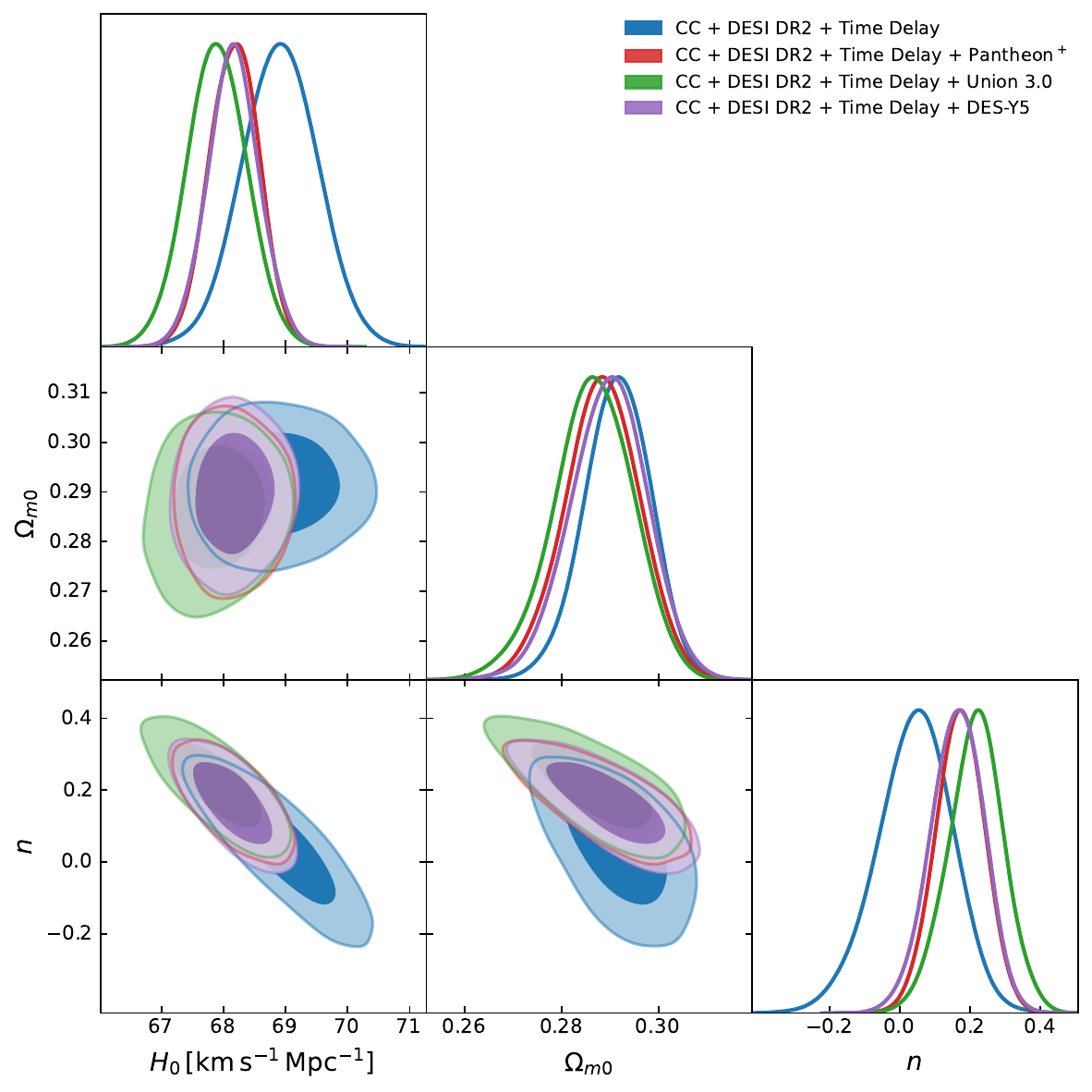}
    \caption{One- and two-dimensional marginalized posterior distributions for the parameters of the normalized power-law $f_{\rm I}(Q)$ model. The contours show the $68\%$ and $95\%$ confidence regions obtained from CC + DESI DR2 + Time-delay and the three combinations that additionally include Pantheon$^+$, Union 3.0, or DES Y5.}
    \label{fig_contour_fQI}
\end{figure}

The corresponding one- and two-dimensional marginalized posterior distributions are shown in Fig.~\ref{fig_contour_fQI}. The contours are well localized for all four dataset combinations, with the inclusion of Type Ia Supernovae providing tighter constraints in comparison with CC + DESI DR2 + Time-delay alone. The strongest correlation is found between the model parameters associated with the modification of the gravitational sector, reflecting the fact that changes in the power-law exponent can be compensated by changes in the corresponding normalization. The posterior distributions of $H_0$ and $\Omega_{m0}$ also show a correlation, although it is less pronounced than the correlation between the modified-gravity parameters. Overall, the contour plots indicate that the parameter space of the normalized power-law model is well constrained by the combined observations.

\begin{table*}
\centering
\caption{Marginalized constraints on the free parameters $H_0$, $\Omega_{m0}$, and $n$ of the normalized power-law $f_{\rm I}(Q)$ model at the $68\%$ confidence level for different combinations of cosmological observations.}
\label{tab:mcmc_constraints_fQI}
\begin{tabular}{lccc}
\hline\hline
Data set & $H_{0} \left[\mathrm{km\,s^{-1}\,Mpc^{-1}}\right]$ & $\Omega_{m0}$ & $n$ \\
\hline
CC + DESI DR2 + Time-delay
& $68.9139^{+0.6063}_{-0.6187}$ 
& $0.2916^{+0.0067}_{-0.0069}$
& $0.0460^{+0.1023}_{-0.1096}$ \\[1.5ex]

CC + DESI DR2 + Time-delay + Pantheon$^+$
& $68.1804^{+0.3909}_{-0.4068}$
& $0.2884^{+0.0076}_{-0.0077}$
& $0.1715^{+0.0693}_{-0.0699}$ \\[1.5ex]

CC + DESI DR2 + Time-delay + Union~3.0
& $67.8902^{+0.4881}_{-0.4704}$
& $0.2866^{+0.0082}_{-0.0081}$
& $0.2182^{+0.0739}_{-0.0813}$ \\[1.5ex]

CC + DESI DR2 + Time-delay + DES~Y5
& $68.1624^{+0.4121}_{-0.4074}$
& $0.2899^{+0.0076}_{-0.0080}$
& $0.1652^{+0.0738}_{-0.0775}$ \\
\hline\hline
\end{tabular}
\end{table*}

\begin{figure}
    \centering
    \includegraphics[width=1\linewidth]{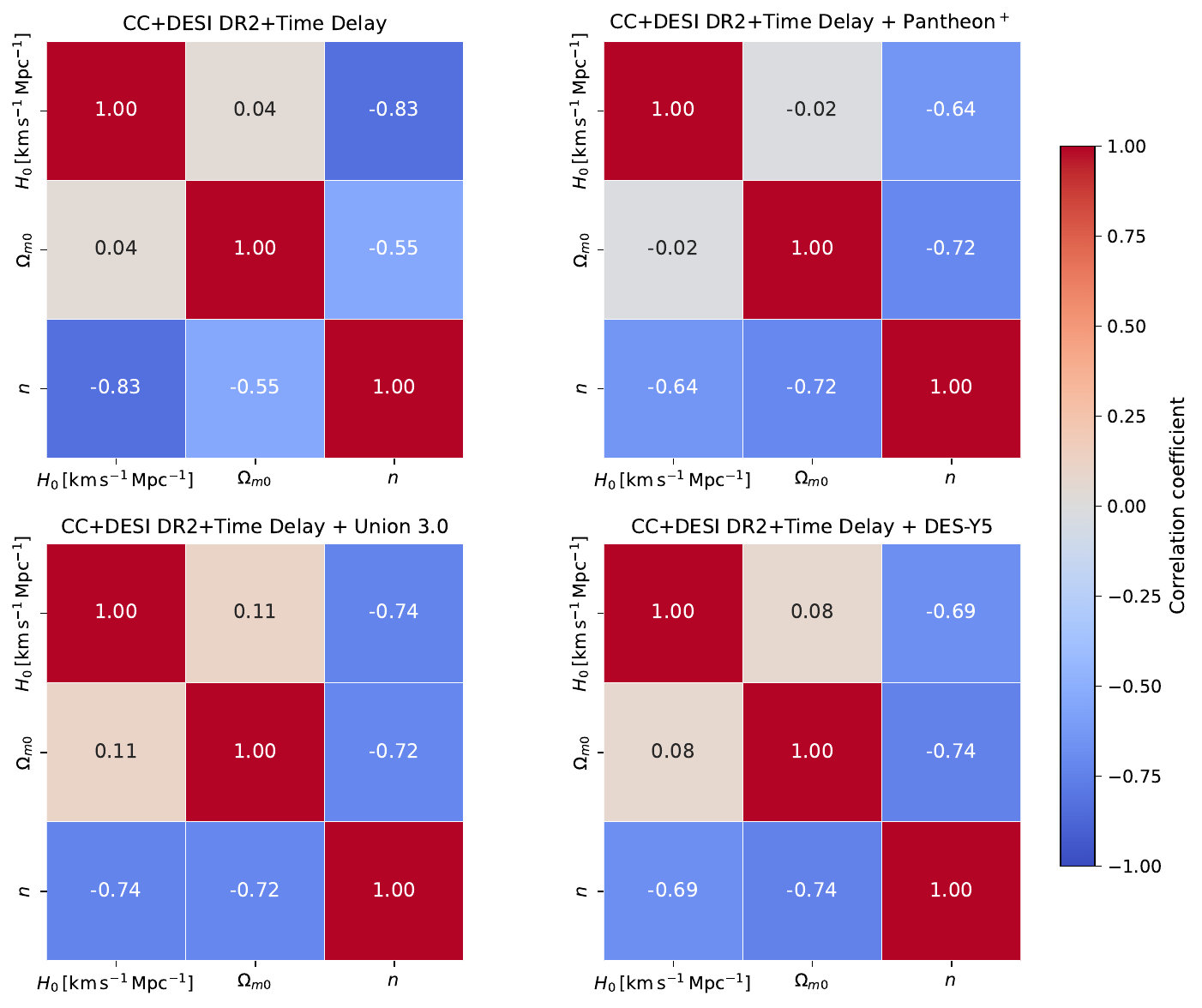}
    \caption{Correlation matrix for the parameters of the normalized power-law $f_{\rm I}(Q)$ model for the different combinations of observational datasets. The matrices illustrate the strength and direction of the correlations among the fitted parameters.}
    \label{fig_fQ1_correlation}
\end{figure}

Figure~\ref{fig_fQ1_correlation} shows the corresponding correlation matrices. The parameters describing the modification of the gravitational sector display the strongest mutual correlation, consistent with the narrow and tilted contours seen in the two-dimensional posterior distributions. The correlations involving $H_0$ and $\Omega_{m0}$ are comparatively weaker, indicating that the combined datasets constrain the background expansion parameters independently of the power-law modification to a reasonable extent.

\subsection{Constraints on the square-root exponential model}\label{sec_results_fQII}

We next consider the square-root exponential form $f_{\rm II}(Q)$. The marginalized constraints on $H_0$, $\Omega_{m0}$, and $m$ are listed in Table~\ref{tab:mcmc_constraints_fQII}. For CC + DESI DR2 + Time-delay, the constraints are $H_0=67.9802^{+0.7757}_{-2.3521}$, $\Omega_{m0}=0.3020^{+0.0165}_{-0.0084}$, and $m=4.0212^{+1.3198}_{-1.3620}$. In comparison with the normalized power-law model, the posterior distributions are more asymmetric, particularly for $H_0$. This behavior becomes more pronounced for some of the supernova combinations, most notably for the DES Y5 dataset.

\begin{figure}
    \centering
    \includegraphics[width=1\linewidth]{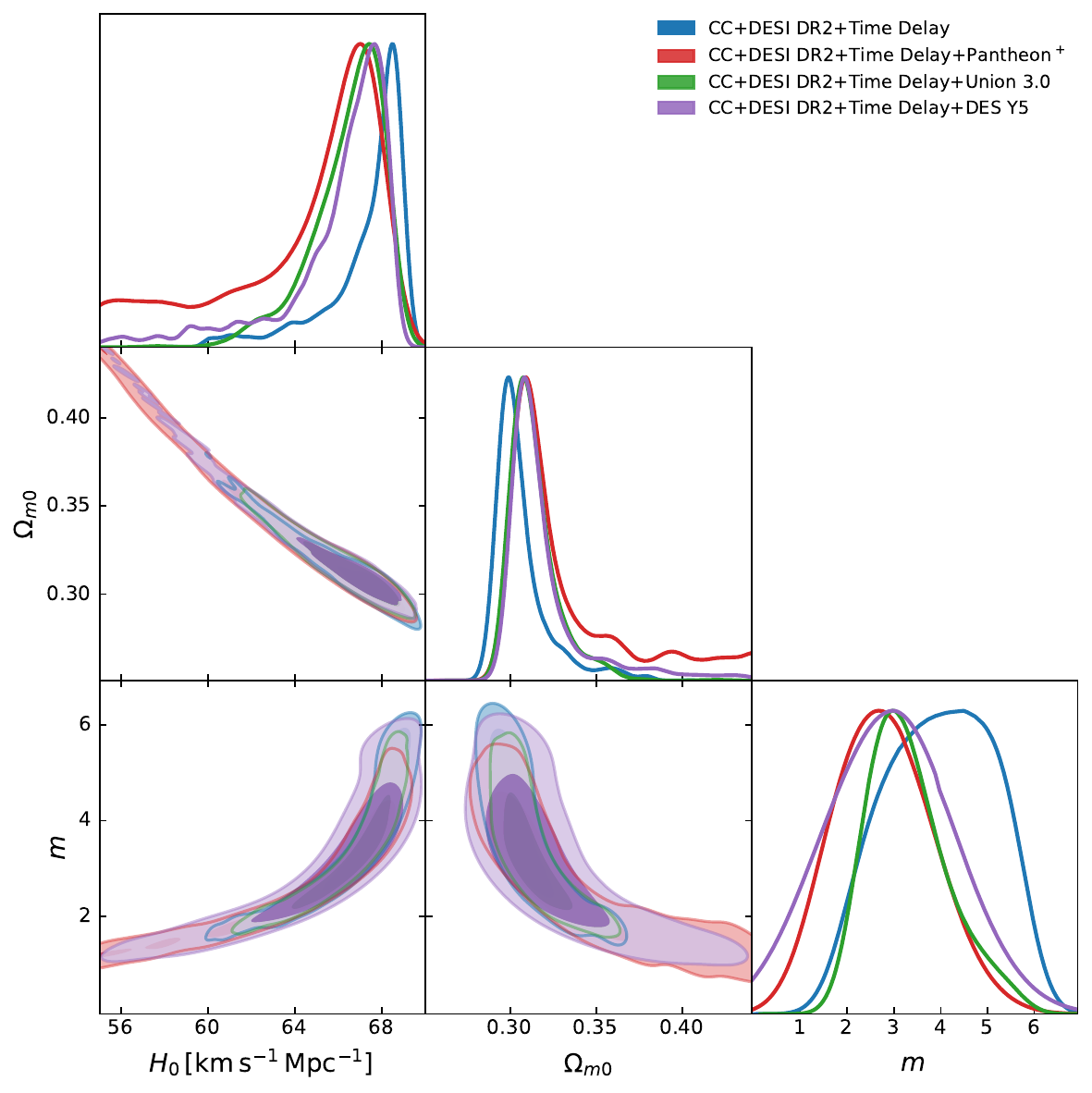}
    \caption{One- and two-dimensional marginalized posterior distributions for the parameters of the square-root exponential $f_{\rm II}(Q)$ model. The contours show the $68\%$ and $95\%$ confidence regions obtained from CC + DESI DR2 + Time-delay and the three combinations that additionally include Pantheon$^+$, Union 3.0, or DES Y5.}
    \label{fig_contour_fQII}
\end{figure}

The addition of Type Ia Supernovae shifts the preferred matter density towards slightly larger values. Pantheon$^+$ gives $\Omega_{m0}=0.3166^{+0.0600}_{-0.0130}$, while Union 3.0 and DES Y5 give values close to $0.316$ and $0.315$, respectively. The parameter $m$ remains of order unity to a few across all combinations, with the Pantheon$^+$ and Union 3.0 datasets giving values below the constraint obtained from CC + DESI DR2 + Time-delay. The posterior distributions for this model are shown in Fig.~\ref{fig_contour_fQII}.

\begin{table*}
\centering
\caption{Marginalized constraints on the free parameters $H_0$, $\Omega_{m0}$, and $m$ of the square-root exponential $f_{\rm II}(Q)$ model at the $68\%$ confidence level for different combinations of cosmological observations.}
\label{tab:mcmc_constraints_fQII}
\begin{tabular}{lccc}
\hline\hline
Data set & $H_{0}~\left[\mathrm{km\,s^{-1}\,Mpc^{-1}}\right]$ & $\Omega_{m0}$ & $m$ \\ 
\hline
CC + DESI DR2 + Time-delay
& $67.9802^{+0.7757}_{-2.3521}$
& $0.3020^{+0.0165}_{-0.0084}$
& $4.0212^{+1.3198}_{-1.3620}$ \\[1.5ex]

CC + DESI DR2 + Time-delay + Pantheon$^+$
& $65.8591^{+1.8408}_{-6.0809}$
& $0.3166^{+0.0600}_{-0.0130}$
& $2.7461^{+1.0679}_{-1.1420}$ \\[1.5ex]

CC + DESI DR2 + Time-delay + Union~3.0
& $66.2337^{+1.3488}_{-2.7271}$
& $0.3158^{+0.0224}_{-0.0113}$
& $2.9279^{+0.8213}_{-0.7657}$ \\[1.5ex]

CC + DESI DR2 + Time-delay + DES~Y5
& $66.3544^{+1.5278}_{-9.9508}$
& $0.3147^{+0.1077}_{-0.0113}$
& $2.9731^{+1.1670}_{-1.6735}$ \\
\hline\hline
\end{tabular}
\end{table*}

\begin{figure}
    \centering
    \includegraphics[width=1\linewidth]{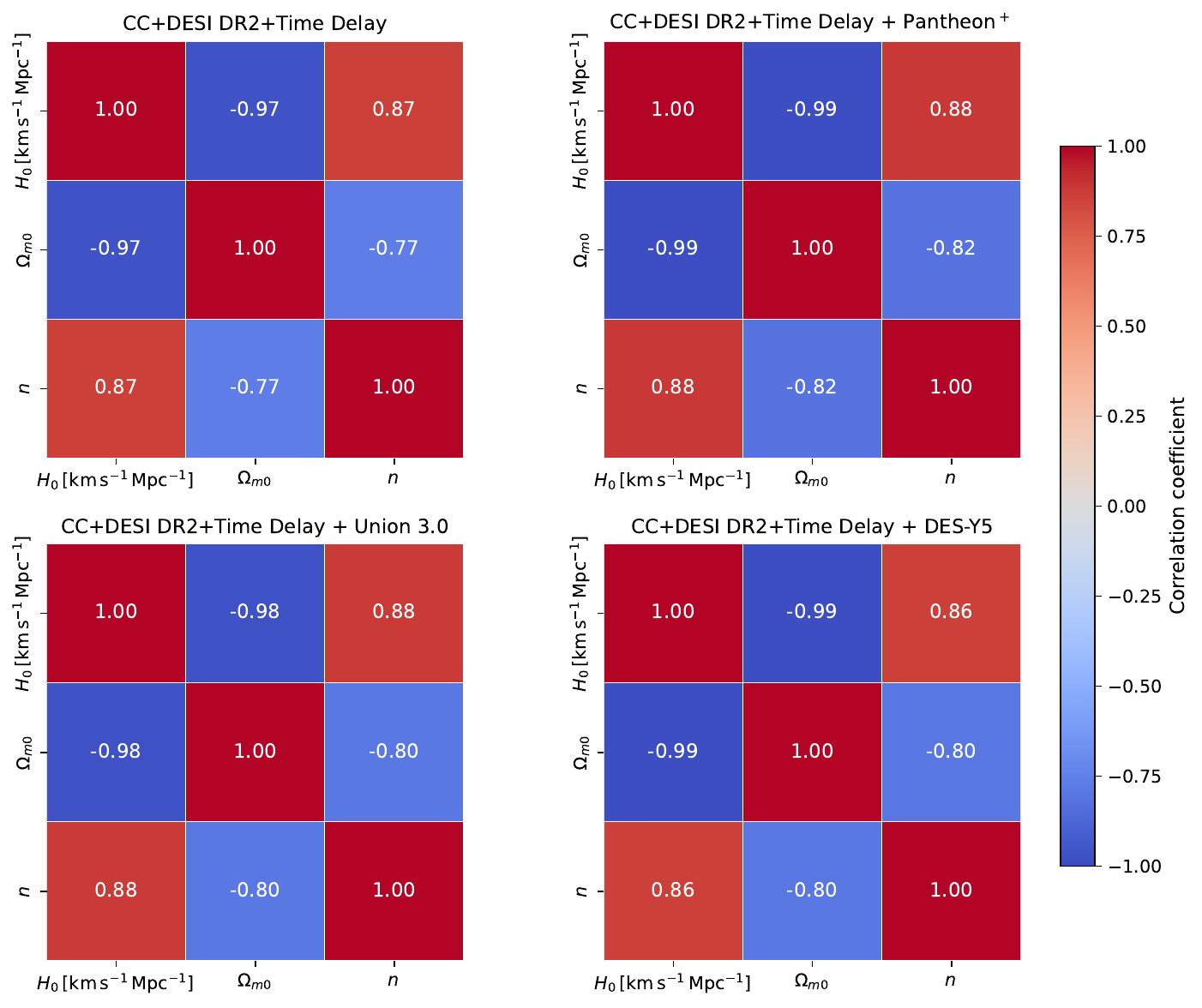}
    \caption{Correlation matrix for the parameters of the square-root exponential $f_{\rm II}(Q)$ model for the different combinations of observational datasets. The matrices show the strength and direction of the correlations among $H_0$, $\Omega_{m0}$, and $m$.}
    \label{fig_fQII_correlation}
\end{figure}

The correlation matrices in Fig.~\ref{fig_fQII_correlation} show that the parameters of the square-root exponential model are correlated to different degrees depending on the dataset combination. These correlations are also reflected in the orientation and extent of the two-dimensional posterior contours. The larger asymmetric uncertainties obtained for some combinations indicate that the present observations leave a broader allowed region for this model than for the normalized power-law form.

\subsection{Statistical comparison with $\Lambda$CDM}\label{sec_model_comparison}

We compare the statistical performance of the two $f(Q)$ models with the flat $\Lambda$CDM model using the minimum chi-square, Akaike Information Criterion (AIC), and Bayesian Information Criterion (BIC). The minimum chi-square, $\chi^2_{\rm min}$, measures the quality of the fit to the observational data. The AIC is defined as
\begin{equation}
\mathrm{AIC}=\chi^2_{\rm min}+2k,
\end{equation}
where $k$ is the number of free parameters. The BIC is given by
\begin{equation}
\mathrm{BIC}=\chi^2_{\rm min}+k\ln N,
\end{equation}
{
where $N$ is the total number of observational data points included in the analysis. In the present work, the four dataset combinations contain $N=52$, $1642$, $74$, and $1687$ data points for CC+DESI DR2+Time-delay, CC+DESI DR2+Time-delay+Pantheon$^+$, CC+DESI DR2+Time-delay+Union 3.0, and CC+DESI DR2+Time-delay+DES Y5, respectively. Since the $f(Q)$ models contain one additional free parameter compared with the flat $\Lambda$CDM model, a lower $\chi^2_{\rm min}$ alone does not necessarily mean that the model provides a better description of the data. The extra parameter gives the model more freedom to fit the observed expansion history, so part of the improvement in $\chi^2_{\rm min}$ may simply result from this increased flexibility rather than from a genuine preference in the data. This raises the possibility of overfitting, where the extra parameter may account for statistical fluctuations in the data rather than a feature that is required by the observations. The AIC and BIC take this additional parameter into account through a penalty for model complexity and therefore help us determine whether the improvement in the fit is large enough to justify the more complex model. This is particularly relevant for the larger datasets, for which the BIC penalty becomes more important. Thus, a lower $\chi^2_{\rm min}$ does not by itself provide evidence in favor of $f(Q)$ gravity; the improvement should remain after the penalty for the additional parameter is taken into account.
}

For a direct comparison, we define
\begin{align}
\Delta\chi^2_{\Lambda\mathrm{CDM}} &= \chi^2_{\rm min}(f(Q))-\chi^2_{\rm min}(\Lambda\mathrm{CDM}),\\
\Delta\mathrm{AIC}_{\Lambda\mathrm{CDM}} &= \mathrm{AIC}_{f(Q)}-\mathrm{AIC}_{\Lambda\mathrm{CDM}},\\
\Delta\mathrm{BIC}_{\Lambda\mathrm{CDM}} &= \mathrm{BIC}_{f(Q)}-\mathrm{BIC}_{\Lambda\mathrm{CDM}}.
\end{align}
Negative values therefore favor the corresponding $f(Q)$ model, whereas positive values favor $\Lambda$CDM. For the information criteria, $|\Delta X|<2$ indicates comparable statistical support, $2\leq|\Delta X|<6$ indicates positive evidence, $6\leq|\Delta X|<10$ indicates strong evidence, and $|\Delta X|\geq10$ indicates very strong evidence.

\begin{table*}
\centering
\caption{Statistical comparison between the normalized power-law $f_{\rm I}(Q)$ model and flat $\Lambda$CDM. The quantities $\Delta\chi^2_{\Lambda\mathrm{CDM}}$, $\Delta\mathrm{AIC}_{\Lambda\mathrm{CDM}}$, and $\Delta\mathrm{BIC}_{\Lambda\mathrm{CDM}}$ are defined as the corresponding $f_{\rm I}(Q)$ values minus the $\Lambda$CDM values. Negative differences favor $f_{\rm I}(Q)$, while positive differences favor $\Lambda$CDM.}
\label{tab:model_comparison_fQI}
\begin{tabular}{lcccccc}
\hline
Dataset & $\chi^2_{\rm min}$ & $\Delta\chi^2_{\Lambda\mathrm{CDM}}$ & AIC & $\Delta\mathrm{AIC}_{\Lambda\mathrm{CDM}}$ & BIC & $\Delta\mathrm{BIC}_{\Lambda\mathrm{CDM}}$ \\
\hline
CC + DESI DR2 + Time Delay
& 41.1948 & $+0.5614$ & 47.1948 & $+2.5614$ & 53.0485 & $+4.5126$ \\[1.5ex]
CC + DESI DR2 + Time Delay + PantheonPlus
& 1456.2751 & $-1.8784$ & 1462.2751 & $+0.1216$ & 1478.4861 & $+5.5253$ \\[1.5ex]
CC + DESI DR2 + Time Delay + Union 3.0
& 72.1039 & $-1.6314$ & 78.1039 & $+0.3686$ & 85.0161 & $+2.6727$ \\[1.5ex]
CC + DESI DR2 + Time Delay + DES-Y5
& 1522.0016 & $-1.1191$ & 1528.0016 & $+0.8809$ & 1544.2937 & $+6.3116$ \\
\hline
\end{tabular}
\end{table*}

Table~\ref{tab:model_comparison_fQI} shows that the normalized power-law model provides a fit that is close to that of $\Lambda$CDM. For CC + DESI DR2 + Time Delay, the minimum chi-square is slightly larger than that of $\Lambda$CDM, with $\Delta\chi^2_{\Lambda\mathrm{CDM}}=+0.56$. Once the supernova data are included, the difference becomes negative for all three compilations. The largest improvement occurs for Pantheon$^+$, although the reduction in chi-square remains modest. Thus, the normalized power-law model can reproduce the observed expansion history with a fit comparable to that of $\Lambda$CDM.

The AIC gives a similar picture. For the three combinations containing Type Ia Supernovae, the AIC differences are all smaller than two in magnitude. This indicates that the additional parameter of $f_{\rm I}(Q)$ does not lead to a substantial loss of statistical support relative to $\Lambda$CDM. The situation is somewhat different for CC + DESI DR2 + Time Delay, where $\Delta\mathrm{AIC}_{\Lambda\mathrm{CDM}}=+2.56$ gives a mild preference for $\Lambda$CDM. The BIC differences are positive for all four combinations. This is expected because the BIC imposes a stronger penalty for the additional parameter, particularly for the larger datasets. Nevertheless, the BIC differences remain moderate for the first three combinations and become larger for DES Y5. These results indicate that $f_{\rm I}(Q)$ is statistically competitive with $\Lambda$CDM, although the information criteria do not provide a clear preference for the additional parameter.

\begin{table*}
\centering
\caption{Statistical comparison between the square-root exponential $f_{\rm II}(Q)$ model and flat $\Lambda$CDM. The quantities $\Delta\chi^2_{\Lambda\mathrm{CDM}}$, $\Delta\mathrm{AIC}_{\Lambda\mathrm{CDM}}$, and $\Delta\mathrm{BIC}_{\Lambda\mathrm{CDM}}$ are defined as the corresponding $f_{\rm II}(Q)$ values minus the $\Lambda$CDM values. Negative differences favor $f_{\rm II}(Q)$, while positive differences favor $\Lambda$CDM.}
\label{tab:model_comparison_fQII}
\begin{tabular}{lcccccc}
\hline
Dataset & $\chi^2_{\rm min}$ & $\Delta\chi^2_{\Lambda\mathrm{CDM}}$ & AIC & $\Delta\mathrm{AIC}_{\Lambda\mathrm{CDM}}$ & BIC & $\Delta\mathrm{BIC}_{\Lambda\mathrm{CDM}}$ \\
\hline
CC + DESI DR2 + Time Delay
& 33.0258 & $-7.6076$ & 39.0258 & $-5.6076$ & 44.8796 & $-3.6563$ \\[1.5ex]
CC + DESI DR2 + Time Delay + PantheonPlus
& 1445.6181 & $-12.5354$ & 1451.6181 & $-10.5354$ & 1467.8291 & $-5.1317$ \\[1.5ex]
CC + DESI DR2 + Time Delay + Union 3.0
& 61.3274 & $-12.4079$ & 67.3274 & $-10.4079$ & 74.2396 & $-8.1038$ \\[1.5ex]
CC + DESI DR2 + Time Delay + DES-Y5
& 1521.7989 & $-1.3218$ & 1527.7989 & $+0.6782$ & 1544.0910 & $+6.1089$ \\
\hline
\end{tabular}
\end{table*}

The statistical behavior of the square-root exponential model is markedly different, as shown in Table~\ref{tab:model_comparison_fQII}. For CC + DESI DR2 + Time Delay, the minimum chi-square is lower than that of $\Lambda$CDM by $7.61$. The improvement remains after accounting for the additional parameter, with $\Delta\mathrm{AIC}_{\Lambda\mathrm{CDM}}=-5.61$ and $\Delta\mathrm{BIC}_{\Lambda\mathrm{CDM}}=-3.66$. Thus, all three measures favor $f_{\rm II}(Q)$ for this dataset combination.

The preference becomes stronger when Pantheon$^+$ is added. The minimum chi-square decreases by $12.54$ relative to $\Lambda$CDM, while the AIC difference reaches $-10.54$. The BIC difference is also negative, with $\Delta\mathrm{BIC}_{\Lambda\mathrm{CDM}}=-5.13$. The improvement in the fit therefore remains after the penalty associated with the additional parameter. According to the AIC, the preference is very strong, while the BIC indicates positive evidence in favor of $f_{\rm II}(Q)$.

The combination with Union 3.0 gives a similar result. The minimum chi-square is lower than the $\Lambda$CDM value by $12.41$, and both information criteria remain negative. In particular, $\Delta\mathrm{AIC}_{\Lambda\mathrm{CDM}}=-10.41$ corresponds to very strong evidence according to the AIC, while $\Delta\mathrm{BIC}_{\Lambda\mathrm{CDM}}=-8.10$ indicates strong evidence according to the BIC. Among the dataset combinations considered here, Union 3.0 provides the largest preference for $f_{\rm II}(Q)$ according to the BIC.

The DES Y5 combination gives a different result. Although the minimum chi-square is still slightly lower than that of $\Lambda$CDM, with $\Delta\chi^2_{\Lambda\mathrm{CDM}}=-1.32$, the AIC difference is close to zero and positive, $\Delta\mathrm{AIC}_{\Lambda\mathrm{CDM}}=+0.68$. The BIC difference is also positive and reaches $+6.11$. Hence, once the additional parameter is taken into account, the information criteria favor $\Lambda$CDM for this particular supernova compilation. This difference between DES Y5 and the other supernova samples indicates that the relative statistical performance of the model depends on the dataset combination.        
    
\begin{figure}
    \centering
    \includegraphics[width=1\linewidth]{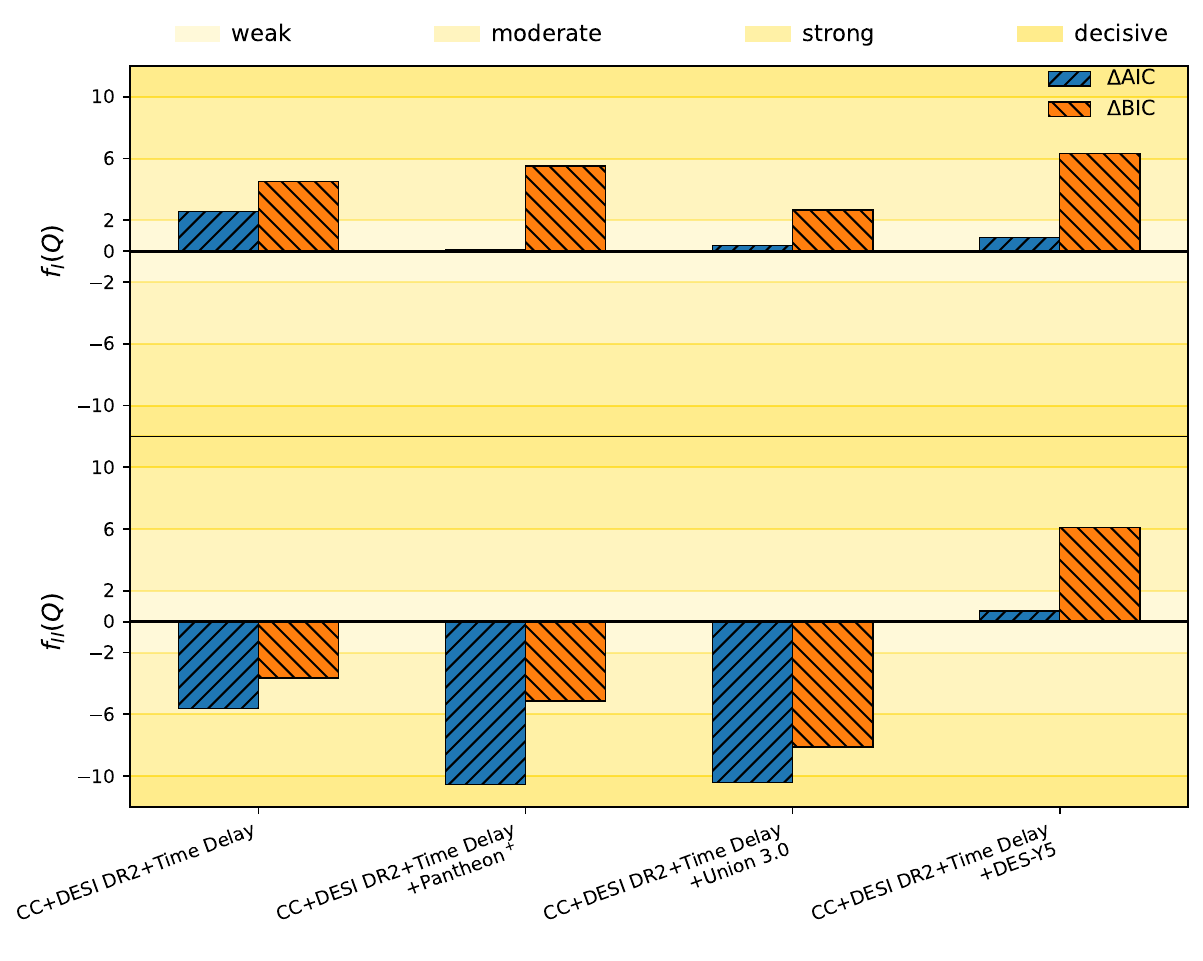}
    \caption{Comparison of the statistical preference of the two $f(Q)$ models relative to flat $\Lambda$CDM. The upper and lower panels correspond to the normalized power-law $f_{\rm I}(Q)$ and square-root exponential $f_{\rm II}(Q)$ models, respectively. The left and right panels show $\Delta\mathrm{AIC}_{\Lambda\mathrm{CDM}}$ and $\Delta\mathrm{BIC}_{\Lambda\mathrm{CDM}}$, respectively. Negative values favor the corresponding $f(Q)$ model, while positive values favor $\Lambda$CDM. The horizontal shaded regions indicate the commonly used ranges of statistical support according to the magnitude of the information-criterion differences.}
    \label{fig_model_comparison}
\end{figure}

Figure~\ref{fig_model_comparison} summarizes the comparison of both $f(Q)$ models with $\Lambda$CDM. The normalized power-law model remains close to the standard cosmological model, with no clear statistical preference from the information criteria. In contrast, the square-root exponential model gives substantially negative AIC and BIC differences for the CC + DESI DR2 + Time Delay, Pantheon$^+$, and Union 3.0 combinations. The DES Y5 combination is the only case in which both information criteria do not favor the modified-gravity model.

Taken together, the statistical comparison clearly distinguishes the two $f(Q)$ models considered in this work. The normalized power-law model remains close to $\Lambda$CDM and provides a comparable fit to the data, although the information criteria do not favor its additional parameter. In contrast, the square-root exponential model gives a substantially better fit for the CC + DESI DR2 + Time Delay, Pantheon$^+$, and Union 3.0 combinations, and this improvement remains after the penalty for the additional parameter is included. The preference for $f_{\rm II}(Q)$ is particularly strong for the Pantheon$^+$ and Union 3.0 datasets, where both AIC and BIC favor the modified-gravity model. The DES Y5 combination does not show the same preference, with both information criteria favoring $\Lambda$CDM. Thus, although the preference depends on the choice of supernova compilation, the overall results show that the square-root exponential $f_{\rm II}(Q)$ model provides a favorable alternative to $\Lambda$CDM for several combinations of current cosmological observations.

%% file: 05_discus_concl.tex
\section{Discussion and Conclusions}\label{sec5}

The late-time accelerated expansion of the Universe remains one of the central problems in modern cosmology. Within the standard $\Lambda$CDM model, the acceleration is attributed to a cosmological constant, although its physical origin is still unknown. Modified theories of gravity provide another possibility in which the observed acceleration can arise from a modification of the gravitational sector itself. In this work, we have investigated this possibility within symmetric teleparallel gravity, where gravity is described through the non-metricity scalar $Q$ rather than spacetime curvature or torsion \cite{BeltranJimenez:2017tkd}. We considered two forms of $f(Q)$ gravity: the normalized power-law model $f_{\rm I}(Q)$ and the square-root exponential model $f_{\rm II}(Q)$. Both models recover the standard symmetric teleparallel equivalent of GR for suitable parameter choices and provide simple extensions of the gravitational sector that can be tested directly against cosmological observations.

We constrained the model parameters using four combinations of current observations: Cosmic Chronometer measurements, DESI DR2 BAO data, strong gravitational lensing time-delay observations, and Type Ia Supernovae data from Pantheon$^+$, Union 3.0, and DES Y5. The resulting constraints allow us to examine whether the additional freedom introduced by $f(Q)$ gravity leads to a statistically meaningful improvement over flat $\Lambda$CDM.

The main results of our analysis can be summarized as follows:     


    (i) The posterior constraints on $H_0$, $\Omega_{m0}$, and the corresponding model parameter are presented in Tables~\ref{tab:mcmc_constraints_fQI} and \ref{tab:mcmc_constraints_fQII} and the corresponding contour plots are shown in Figs.~\ref{fig_contour_fQI} and \ref{fig_contour_fQII}. The inclusion of the Type Ia Supernovae samples generally leads to tighter constraints compared with the combination of CC + DESI DR2 + Time-delay alone. The Pantheon$^+$ combination gives a somewhat different constraint on $H_0$, which indicates the sensitivity of the inferred parameter values to the supernova compilation.

   (ii) The posterior distributions show a strong correlation between the parameters of the modified-gravity sector. In particular, the parameter combinations entering the normalized power-law model are strongly correlated, reflecting the fact that different combinations of these parameters can produce similar expansion histories. The correlation matrices shown in Figs.~\ref{fig_fQ1_correlation} and \ref{fig_fQII_correlation} provide a direct view of these parameter degeneracies. The square-root exponential model also shows parameter correlations, although their structure differs from that of the power-law model because of its different dependence on $H$.

    (iii) An important result comes from the statistical comparison with flat $\Lambda$CDM. For $f_{\rm I}(Q)$, the values of $\Delta\chi^2_{\Lambda\mathrm{CDM}}$ are slightly negative for the three combinations containing Type Ia Supernovae data, indicating a small improvement in the minimum $\chi^2$. However, this improvement is not large enough to overcome the penalty associated with the additional parameter in all of the information criteria. As shown in Table~\ref{tab:model_comparison_fQI}, the AIC differences for these three combinations remain smaller than two, while the BIC still gives a mild preference for $\Lambda$CDM. Thus, the normalized power-law model provides a good description of the observations, but the present data do not give strong statistical support for its additional degree of freedom.

    (iv) The square-root exponential model gives a more interesting result. As shown in Table~\ref{tab:model_comparison_fQII}, $f_{\rm II}(Q)$ produces a substantial reduction in $\chi^2_{\rm min}$ for the CC + DESI DR2 + Time-delay, CC + DESI DR2 + Time-delay + Pantheon$^+$, and CC + DESI DR2 + Time-delay + Union 3.0 combinations. The improvement remains after the parameter penalty is included in both AIC and BIC. In particular, the Pantheon$^+$ and Union 3.0 combinations give negative values of $\Delta\mathrm{AIC}$ and $\Delta\mathrm{BIC}$, showing that the better fit provided by $f_{\rm II}(Q)$ is not simply a consequence of introducing an additional free parameter.

    (v) The DES Y5 combination gives a different result. In this case, the reduction in $\chi^2_{\rm min}$ for $f_{\rm II}(Q)$ is relatively small, while both AIC and BIC favor $\Lambda$CDM. This difference between the supernova compilations is important because it shows that the preference for modified gravity is not independent of the observational dataset. Nevertheless, three of the four combinations considered here give a statistically favorable result for $f_{\rm II}(Q)$, with the strongest preference obtained when Pantheon$^+$ and Union 3.0 are combined with CC, DESI DR2, and time-delay data.

    (vi) The comparison of the two models is illustrated in Fig.~\ref{fig_model_comparison}. The results show a clear distinction between the two functional forms. The normalized power-law model remains statistically close to $\Lambda$CDM, whereas the square-root exponential model gives a noticeably better fit for several combinations of observations. In particular, the negative values of $\Delta\mathrm{AIC}$ and $\Delta\mathrm{BIC}$ for the Pantheon$^+$ and Union 3.0 combinations indicate that the improvement in the fit is sufficient to compensate for the additional model parameter. This makes $f_{\rm II}(Q)$ the more favorable of the two models considered in the present analysis.

\begin{figure} 
\centering 
\includegraphics[width=1\linewidth]{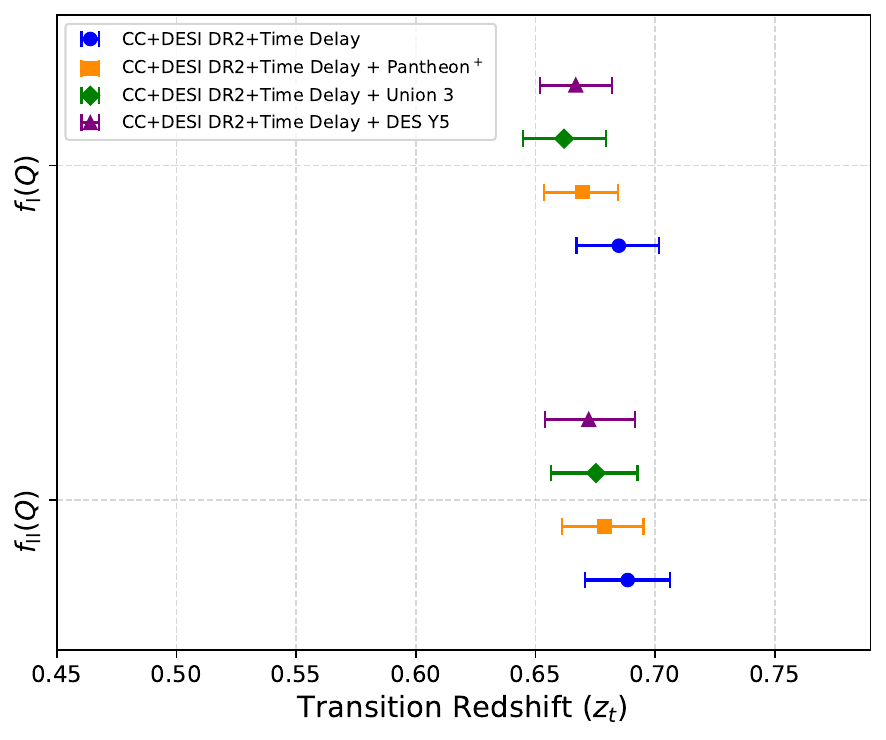} 
\caption{Transition redshift values for different datasets across two dark energy equations of state. The plot shows error bars for various datasets, including Joint (corresponding to CC+DESI DR2+Time-delay), Joint + Pantheon Plus, Joint + Union3.0, and Joint + DES Y5. Additionally, bands for the SH0ES Pantheon sample (with only statistical errors) and Planck-2018 ($\Lambda$CDM) are highlighted.} 
\label{fig_transi_redshift} 
\end{figure}

We also examine the transition from cosmic deceleration to acceleration. The transition redshift $z_{\rm t}$ is determined from the condition $q(z_{\rm t})=0$, where $q$ is the deceleration parameter. The resulting values are listed in Table~\ref{tab:zt_fQ} and shown in Fig.~\ref{fig_transi_redshift}. For the CC + DESI DR2 + Time-delay combination, we obtain $z_{\rm t}=0.6849^{+0.0167}_{-0.0177}$ and $z_{\rm t}=0.6886^{+0.0178}_{-0.0178}$ for $f_{\rm I}(Q)$ and $f_{\rm II}(Q)$, respectively. The inclusion of Type Ia Supernovae shifts the transition to slightly lower redshifts. For $f_{\rm I}(Q)$, the transition redshift reaches its lowest value for the Union 3.0 combination, while the values obtained from $f_{\rm II}(Q)$ remain within a relatively narrow range. Overall, both models give $z_{\rm t}$ in the range $0.66 \leq z_{\rm t} \leq 0.69$ across all four dataset combinations.

\begin{table} 
\centering 
\renewcommand{\arraystretch}{1.2} 
\caption{Transition redshift $z_{\rm t}$ obtained from the condition $q(z_{\rm t})=0$ for the normalized power-law $f_{\rm I}(Q)$ and square-root exponential $f_{\rm II}(Q)$ models using different combinations of observational datasets. The quoted asymmetric uncertainties correspond to the $68\%$ confidence level.}
\label{tab:zt_fQ} 
\begin{tabular}{lcc} \hline\hline 
Dataset & \multicolumn{2}{c}{Transition Redshift $\left(z_\mathrm{t}\right)$} \\ \cline{2-3} & $f_{\rm I}(Q)$ & $f_{\rm II}(Q)$ \\ \hline 
CC + DESI DR2 + Time Delay & $0.6849_{-0.0177}^{+0.0167}$ & $0.6886_{-0.0178}^{+0.0178}$ \\[1.5ex] 
CC + DESI DR2 + Time Delay + Pantheon$^+$ & $0.6697_{-0.0161}^{+0.0149}$ & $0.6789_{-0.0178}^{+0.0163}$ \\[1.5ex] 
CC + DESI DR2 + Time Delay + Union3 & $0.6620_{-0.0170}^{+0.0177}$ & $0.6753_{-0.0186}^{+0.0174}$ \\[1.5ex] 
CC + DESI DR2 + Time Delay + DESY5 & $0.6669_{-0.0149}^{+0.0151}$ & $0.6723_{-0.0182}^{+0.0192}$ \\ \hline\hline 
\end{tabular} 
\end{table}

The close agreement between the two models in the transition redshift is noteworthy. Although the square-root exponential model is statistically preferred for several dataset combinations, this difference does not lead to a large change in the predicted epoch of the deceleration-to-acceleration transition. The values shown in Fig.~\ref{fig_transi_redshift} are also consistent with the range inferred from late-time Type Ia Supernova observations and remain close to the transition expected in the standard $\Lambda$CDM cosmology. Thus, both functional forms provide a consistent description of the onset of the late-time accelerated phase.

An important point is that the statistical preference for $f_{\rm II}(Q)$ does not come at the cost of an unusual expansion history. The model gives a transition epoch similar to that obtained from $f_{\rm I}(Q)$ and $\Lambda$CDM, while providing a lower $\chi^2_{\rm min}$ for several combinations of current observations. This combination of a conventional transition redshift and an improved statistical fit makes the square-root exponential form particularly interesting for further study.

The results of this work can therefore be summarized in the following three main points:


    I. \textit{Both $f(Q)$ models provide a good description of the observed late-time expansion history.} The normalized power-law model remains close to $\Lambda$CDM, while the square-root exponential model gives a larger improvement in the fit for several dataset combinations.

    II. \textit{The square-root exponential model is statistically favored over $\Lambda$CDM for several important combinations of observations.} In particular, the Pantheon$^+$ and Union 3.0 combinations give negative $\Delta\chi^2$, $\Delta\mathrm{AIC}$, and $\Delta\mathrm{BIC}$, showing that the improvement persists even after the additional model parameter is taken into account. The DES Y5 combination is the only case in which the information criteria favor $\Lambda$CDM.

    III. \textit{Both models predict a similar transition from deceleration to acceleration}, with $z_{\rm t}$ remaining around $0.66$--$0.69$ for all dataset combinations. The agreement of the transition redshift with late-time observational estimates provides an additional consistency check on the background evolution.


The present results suggest that the functional form of $f(Q)$ plays an important role in determining how strongly modified gravity is supported by the data. The normalized power-law model remains a viable alternative to $\Lambda$CDM, although its additional parameter is not strongly favored by the information criteria. The square-root exponential model gives a more favorable statistical result and, in particular, remains preferred after the parameter penalty for the Pantheon$^+$ and Union 3.0 combinations. Therefore, \textit{the present analysis provides evidence that a suitable form of $f(Q)$ gravity can describe the observed late-time acceleration without requiring a cosmological constant as the source of cosmic acceleration.}

At the same time, the results should not be interpreted as evidence that $f(Q)$ gravity is universally preferred over $\Lambda$CDM. The DES Y5 combination gives a preference for the standard model, showing that the statistical conclusion depends to some extent on the choice of supernova compilation. A larger and more homogeneous set of low- and high-redshift observations will therefore be important for determining whether the preference for the square-root exponential model persists.

Finally, the analysis presented here is based on the background expansion history. A complete assessment of $f(Q)$ gravity should also consider the evolution of cosmological perturbations, structure formation, and other observational probes beyond the background level. Such an analysis would provide an important test of whether the favorable background-level results obtained here remain valid when the full cosmological dynamics are considered. We leave this investigation for future work.

%% file: acknowledgements.tex
Kumar, D. is supported by the Henan Provincial Natural Science Foundation under Grant No. 262300421843 and by the Startup Research Fund of the Henan Academy of Sciences under Grant No. 241841219. 
Ray, S., Rani, N. and Dhankar, P.K. would like to acknowledge Inter-University Centre for Astronomy and Astrophysics (IUCAA), Pune, India for providing them Visiting Associateship under which a part of this work was carried out. 
Zhang, F. is supported by the National Natural Science Foundation of China under Grant No. 12305075, the Startup Research Fund of Henan Academy of Sciences under Grant No. 241841223, and Joint Fund for Scientific and Technological Research of Henan Province under Grant No. 235200810101.      

%% file: main_ref.bib
@article{einstein1916foundation,
    author = "Einstein, Albert",
    editor = "Hsu, Jong-Ping and Fine, D.",
    title = "{The foundation of the general theory of relativity.}",
    doi = "10.1002/andp.19163540702",
    journal = "Annalen Phys.",
    volume = "49",
    number = "7",
    pages = "769--822",
    year = "1916"
}

@incollection{einstein1922general,
  title={The general theory of relativity},
  author={Einstein, Albert},
  booktitle={The meaning of relativity},
  pages={54--75},
  year={1922},
  publisher={Springer}
}

@ARTICLE{2025JCAP...02..021A,
       author = {{Adame}, A.~G. and {Aguilar}, J. and {Ahlen}, S. and {Alam}, S. and {Alexander}, D.~M. and {Alvarez}, M. and {Alves}, O. and {Anand}, A. and {Andrade}, U. and {Armengaud}, E. and {Avila}, S. and {Aviles}, A. and {Awan}, H. and {Bahr-Kalus}, B. and {Bailey}, S. and {Baltay}, C. and {Bault}, A. and {Behera}, J. and {BenZvi}, S. and {Bera}, A. and {Beutler}, F. and {Bianchi}, D. and {Blake}, C. and {Blum}, R. and {Brieden}, S. and {Brodzeller}, A. and {Brooks}, D. and {Buckley-Geer}, E. and {Burtin}, E. and {Calderon}, R. and {Canning}, R. and {Carnero Rosell}, A. and {Cereskaite}, R. and {Cervantes-Cota}, J.~L. and {Chabanier}, S. and {Chaussidon}, E. and {Chaves-Montero}, J. and {Chen}, S. and {Chen}, X. and {Claybaugh}, T. and {Cole}, S. and {Cuceu}, A. and {Davis}, T.~M. and {Dawson}, K. and {de la Macorra}, A. and {de Mattia}, A. and {Deiosso}, N. and {Dey}, A. and {Dey}, B. and {Ding}, Z. and {Doel}, P. and {Edelstein}, J. and {Eftekharzadeh}, S. and {Eisenstein}, D.~J. and {Elliott}, A. and {Fagrelius}, P. and {Fanning}, K. and {Ferraro}, S. and {Ereza}, J. and {Findlay}, N. and {Flaugher}, B. and {Font-Ribera}, A. and {Forero-S{\'a}nchez}, D. and {Forero-Romero}, J.~E. and {Frenk}, C.~S. and {Garcia-Quintero}, C. and {Gazta{\~n}aga}, E. and {Gil-Mar{\'\i}n}, H. and {Gontcho a Gontcho}, S. and {Gonzalez-Morales}, A.~X. and {Gonzalez-Perez}, V. and {Gordon}, C. and {Green}, D. and {Gruen}, D. and {Gsponer}, R. and {Gutierrez}, G. and {Guy}, J. and {Hadzhiyska}, B. and {Hahn}, C. and {Hanif}, M.~M.~S. and {Herrera-Alcantar}, H.~K. and {Honscheid}, K. and {Howlett}, C. and {Huterer}, D. and {Ir{\v{s}}i{\v{c}}}, V. and {Ishak}, M. and {Juneau}, S. and {Kara{\c{c}}ayl{\i}}, N.~G. and {Kehoe}, R. and {Kent}, S. and {Kirkby}, D. and {Kremin}, A. and {Krolewski}, A. and {Lai}, Y. and {Lan}, T. -W. and {Landriau}, M. and {Lang}, D. and {Lasker}, J. and {Le Goff}, J.~M. and {Le Guillou}, L. and {Leauthaud}, A. and {Levi}, M.~E. and {Li}, T.~S. and {Linder}, E. and {Lodha}, K. and {Magneville}, C. and {Manera}, M. and {Margala}, D. and {Martini}, P. and {Maus}, M. and {McDonald}, P. and {Medina-Varela}, L. and {Meisner}, A. and {Mena-Fern{\'a}ndez}, J. and {Miquel}, R. and {Moon}, J. and {Moore}, S. and {Moustakas}, J. and {Mueller}, E. and {Mu{\~n}oz-Guti{\'e}rrez}, A. and {Myers}, A.~D. and {Nadathur}, S. and {Napolitano}, L. and {Neveux}, R. and {Newman}, J.~A. and {Nguyen}, N.~M. and {Nie}, J. and {Niz}, G. and {Noriega}, H.~E. and {Padmanabhan}, N. and {Paillas}, E. and {Palanque-Delabrouille}, N. and {Pan}, J. and {Penmetsa}, S. and {Percival}, W.~J. and {Pieri}, M.~M. and {Pinon}, M. and {Poppett}, C. and {Porredon}, A. and {Prada}, F. and {P{\'e}rez-Fern{\'a}ndez}, A. and {P{\'e}rez-R{\`a}fols}, I. and {Rabinowitz}, D. and {Raichoor}, A. and {Ram{\'\i}rez-P{\'e}rez}, C. and {Ramirez-Solano}, S. and {Rashkovetskyi}, M. and {Ravoux}, C. and {Rezaie}, M. and {Rich}, J. and {Rocher}, A. and {Rockosi}, C. and {Roe}, N.~A. and {Rosado-Marin}, A. and {Ross}, A.~J. and {Rossi}, G. and {Ruggeri}, R. and {Ruhlmann-Kleider}, V. and {Samushia}, L. and {Sanchez}, E. and {Saulder}, C. and {Schlafly}, E.~F. and {Schlegel}, D. and {Schubnell}, M. and {Seo}, H. and {Shafieloo}, A. and {Sharples}, R. and {Silber}, J. and {Slosar}, A. and {Smith}, A. and {Sprayberry}, D. and {Tan}, T. and {Tarl{\'e}}, G. and {Taylor}, P. and {Trusov}, S. and {Ure{\~n}a-L{\'o}pez}, L.~A. and {Vaisakh}, R. and {Valcin}, D. and {Valdes}, F. and {Vargas-Maga{\~n}a}, M. and {Verde}, L. and {Walther}, M. and {Wang}, B. and {Wang}, M.~S. and {Weaver}, B.~A. and {Weaverdyck}, N. and {Wechsler}, R.~H. and {Weinberg}, D.~H. and {White}, M. and {Yu}, J. and {Yu}, Y. and {Yuan}, S. and {Y{\`e}che}, C. and {Zaborowski}, E.~A. and {Zarrouk}, P. and {Zhang}, H. and {Zhao}, C. and {Zhao}, R. and {Zhou}, R. and {Zhuang}, T.},
        title = "{DESI 2024 VI: cosmological constraints from the measurements of baryon acoustic oscillations}",
      journal = {\jcap},
         year = 2025,
        month = feb,
       volume = {2025},
       number = {2},
          eid = {021},
        pages = {021},
          doi = {10.1088/1475-7516/2025/02/021},
archivePrefix = {arXiv},
       eprint = {2404.03002},
 primaryClass = {astro-ph.CO},
       adsurl = {https://ui.adsabs.harvard.edu/abs/2025JCAP...02..021A}
}

@ARTICLE{pantheon,
       author = {{Scolnic}, D.~M. and {Jones}, D.~O. and {Rest}, A. and {Pan}, Y.~C. and {Chornock}, R. and {Foley}, R.~J. and {Huber}, M.~E. and {Kessler}, R. and {Narayan}, G. and {Riess}, A.~G. and {Rodney}, S. and {Berger}, E. and {Brout}, D.~J. and {Challis}, P.~J. and {Drout}, M. and {Finkbeiner}, D. and {Lunnan}, R. and {Kirshner}, R.~P. and {Sanders}, N.~E. and {Schlafly}, E. and {Smartt}, S. and {Stubbs}, C.~W. and {Tonry}, J. and {Wood-Vasey}, W.~M. and {Foley}, M. and {Hand}, J. and {Johnson}, E. and {Burgett}, W.~S. and {Chambers}, K.~C. and {Draper}, P.~W. and {Hodapp}, K.~W. and {Kaiser}, N. and {Kudritzki}, R.~P. and {Magnier}, E.~A. and {Metcalfe}, N. and {Bresolin}, F. and {Gall}, E. and {Kotak}, R. and {McCrum}, M. and {Smith}, K.~W.},
        title = "{The Complete Light-curve Sample of Spectroscopically Confirmed SNe Ia from Pan-STARRS1 and Cosmological Constraints from the Combined Pantheon Sample}",
      journal = {\apj},
         year = 2018,
        month = jun,
       volume = {859},
       number = {2},
          eid = {101},
        pages = {101},
          doi = {10.3847/1538-4357/aab9bb},
archivePrefix = {arXiv},
       eprint = {1710.00845},
 primaryClass = {astro-ph.CO},
       adsurl = {https://ui.adsabs.harvard.edu/abs/2018ApJ...859..101S}
}

@ARTICLE{1998AJ....116.1009R,
       author = {{Riess}, Adam G. and {Filippenko}, Alexei V. and {Challis}, Peter and {Clocchiatti}, Alejandro and {Diercks}, Alan and {Garnavich}, Peter M. and {Gilliland}, Ron L. and {Hogan}, Craig J. and {Jha}, Saurabh and {Kirshner}, Robert P. and {Leibundgut}, B. and {Phillips}, M.~M. and {Reiss}, David and {Schmidt}, Brian P. and {Schommer}, Robert A. and {Smith}, R. Chris and {Spyromilio}, J. and {Stubbs}, Christopher and {Suntzeff}, Nicholas B. and {Tonry}, John},
        title = "{Observational Evidence from Supernovae for an Accelerating Universe and a Cosmological Constant}",
      journal = {\aj},
         year = 1998,
        month = sep,
       volume = {116},
       number = {3},
        pages = {1009-1038},
          doi = {10.1086/300499},
archivePrefix = {arXiv},
       eprint = {astro-ph/9805201},
 primaryClass = {astro-ph},
       adsurl = {https://ui.adsabs.harvard.edu/abs/1998AJ....116.1009R}
}

@ARTICLE{1999ApJ...517..565P,
       author = {{Perlmutter}, S. and {Aldering}, G. and {Goldhaber}, G. and {Knop}, R.~A. and {Nugent}, P. and {Castro}, P.~G. and {Deustua}, S. and {Fabbro}, S. and {Goobar}, A. and {Groom}, D.~E. and {Hook}, I.~M. and {Kim}, A.~G. and {Kim}, M.~Y. and {Lee}, J.~C. and {Nunes}, N.~J. and {Pain}, R. and {Pennypacker}, C.~R. and {Quimby}, R. and {Lidman}, C. and {Ellis}, R.~S. and {Irwin}, M. and {McMahon}, R.~G. and {Ruiz-Lapuente}, P. and {Walton}, N. and {Schaefer}, B. and {Boyle}, B.~J. and {Filippenko}, A.~V. and {Matheson}, T. and {Fruchter}, A.~S. and {Panagia}, N. and {Newberg}, H.~J.~M. and {Couch}, W.~J. and {Project}, The Supernova Cosmology},
        title = "{Measurements of {\ensuremath{\Omega}} and {\ensuremath{\Lambda}} from 42 High-Redshift Supernovae}",
      journal = {\apj},
         year = 1999,
        month = jun,
       volume = {517},
       number = {2},
        pages = {565-586},
          doi = {10.1086/307221},
archivePrefix = {arXiv},
       eprint = {astro-ph/9812133},
 primaryClass = {astro-ph},
       adsurl = {https://ui.adsabs.harvard.edu/abs/1999ApJ...517..565P}
}

@ARTICLE{1807.06209,
       author = {{Planck Collaboration} and {Aghanim}, N. and {Akrami}, Y. and {Ashdown}, M. and {Aumont}, J. and {Baccigalupi}, C. and {Ballardini}, M. and {Banday}, A.~J. and {Barreiro}, R.~B. and {Bartolo}, N. and {Basak}, S. and {Battye}, R. and {Benabed}, K. and {Bernard}, J. -P. and {Bersanelli}, M. and {Bielewicz}, P. and {Bock}, J.~J. and {Bond}, J.~R. and {Borrill}, J. and {Bouchet}, F.~R. and {Boulanger}, F. and {Bucher}, M. and {Burigana}, C. and {Butler}, R.~C. and {Calabrese}, E. and {Cardoso}, J. -F. and {Carron}, J. and {Challinor}, A. and {Chiang}, H.~C. and {Chluba}, J. and {Colombo}, L.~P.~L. and {Combet}, C. and {Contreras}, D. and {Crill}, B.~P. and {Cuttaia}, F. and {de Bernardis}, P. and {de Zotti}, G. and {Delabrouille}, J. and {Delouis}, J. -M. and {Di Valentino}, E. and {Diego}, J.~M. and {Dor{\'e}}, O. and {Douspis}, M. and {Ducout}, A. and {Dupac}, X. and {Dusini}, S. and {Efstathiou}, G. and {Elsner}, F. and {En{\ss}lin}, T.~A. and {Eriksen}, H.~K. and {Fantaye}, Y. and {Farhang}, M. and {Fergusson}, J. and {Fernandez-Cobos}, R. and {Finelli}, F. and {Forastieri}, F. and {Frailis}, M. and {Fraisse}, A.~A. and {Franceschi}, E. and {Frolov}, A. and {Galeotta}, S. and {Galli}, S. and {Ganga}, K. and {G{\'e}nova-Santos}, R.~T. and {Gerbino}, M. and {Ghosh}, T. and {Gonz{\'a}lez-Nuevo}, J. and {G{\'o}rski}, K.~M. and {Gratton}, S. and {Gruppuso}, A. and {Gudmundsson}, J.~E. and {Hamann}, J. and {Handley}, W. and {Hansen}, F.~K. and {Herranz}, D. and {Hildebrandt}, S.~R. and {Hivon}, E. and {Huang}, Z. and {Jaffe}, A.~H. and {Jones}, W.~C. and {Karakci}, A. and {Keih{\"a}nen}, E. and {Keskitalo}, R. and {Kiiveri}, K. and {Kim}, J. and {Kisner}, T.~S. and {Knox}, L. and {Krachmalnicoff}, N. and {Kunz}, M. and {Kurki-Suonio}, H. and {Lagache}, G. and {Lamarre}, J. -M. and {Lasenby}, A. and {Lattanzi}, M. and {Lawrence}, C.~R. and {Le Jeune}, M. and {Lemos}, P. and {Lesgourgues}, J. and {Levrier}, F. and {Lewis}, A. and {Liguori}, M. and {Lilje}, P.~B. and {Lilley}, M. and {Lindholm}, V. and {L{\'o}pez-Caniego}, M. and {Lubin}, P.~M. and {Ma}, Y. -Z. and {Mac{\'\i}as-P{\'e}rez}, J.~F. and {Maggio}, G. and {Maino}, D. and {Mandolesi}, N. and {Mangilli}, A. and {Marcos-Caballero}, A. and {Maris}, M. and {Martin}, P.~G. and {Martinelli}, M. and {Mart{\'\i}nez-Gonz{\'a}lez}, E. and {Matarrese}, S. and {Mauri}, N. and {McEwen}, J.~D. and {Meinhold}, P.~R. and {Melchiorri}, A. and {Mennella}, A. and {Migliaccio}, M. and {Millea}, M. and {Mitra}, S. and {Miville-Desch{\^e}nes}, M. -A. and {Molinari}, D. and {Montier}, L. and {Morgante}, G. and {Moss}, A. and {Natoli}, P. and {N{\o}rgaard-Nielsen}, H.~U. and {Pagano}, L. and {Paoletti}, D. and {Partridge}, B. and {Patanchon}, G. and {Peiris}, H.~V. and {Perrotta}, F. and {Pettorino}, V. and {Piacentini}, F. and {Polastri}, L. and {Polenta}, G. and {Puget}, J. -L. and {Rachen}, J.~P. and {Reinecke}, M. and {Remazeilles}, M. and {Renzi}, A. and {Rocha}, G. and {Rosset}, C. and {Roudier}, G. and {Rubi{\~n}o-Mart{\'\i}n}, J.~A. and {Ruiz-Granados}, B. and {Salvati}, L. and {Sandri}, M. and {Savelainen}, M. and {Scott}, D. and {Shellard}, E.~P.~S. and {Sirignano}, C. and {Sirri}, G. and {Spencer}, L.~D. and {Sunyaev}, R. and {Suur-Uski}, A. -S. and {Tauber}, J.~A. and {Tavagnacco}, D. and {Tenti}, M. and {Toffolatti}, L. and {Tomasi}, M. and {Trombetti}, T. and {Valenziano}, L. and {Valiviita}, J. and {Van Tent}, B. and {Vibert}, L. and {Vielva}, P. and {Villa}, F. and {Vittorio}, N. and {Wandelt}, B.~D. and {Wehus}, I.~K. and {White}, M. and {White}, S.~D.~M. and {Zacchei}, A. and {Zonca}, A.},
        title = "{Planck 2018 results. VI. Cosmological parameters}",
      journal = {\aap},
         year = 2020,
        month = sep,
       volume = {641},
          eid = {A6},
        pages = {A6},
          doi = {10.1051/0004-6361/201833910},
archivePrefix = {arXiv},
       eprint = {1807.06209},
 primaryClass = {astro-ph.CO},
       adsurl = {https://ui.adsabs.harvard.edu/abs/2020A&A...641A...6P}
}

@ARTICLE{2011ApJS..192...18K,
       author = {{Komatsu}, E. and {Smith}, K.~M. and {Dunkley}, J. and {Bennett}, C.~L. and {Gold}, B. and {Hinshaw}, G. and {Jarosik}, N. and {Larson}, D. and {Nolta}, M.~R. and {Page}, L. and {Spergel}, D.~N. and {Halpern}, M. and {Hill}, R.~S. and {Kogut}, A. and {Limon}, M. and {Meyer}, S.~S. and {Odegard}, N. and {Tucker}, G.~S. and {Weiland}, J.~L. and {Wollack}, E. and {Wright}, E.~L.},
        title = "{Seven-year Wilkinson Microwave Anisotropy Probe (WMAP) Observations: Cosmological Interpretation}",
      journal = {\apjs},
         year = 2011,
        month = feb,
       volume = {192},
       number = {2},
          eid = {18},
        pages = {18},
          doi = {10.1088/0067-0049/192/2/18},
archivePrefix = {arXiv},
       eprint = {1001.4538},
 primaryClass = {astro-ph.CO},
       adsurl = {https://ui.adsabs.harvard.edu/abs/2011ApJS..192...18K}
}

@ARTICLE{2017MNRAS.465.4895W,
       author = {{Wong}, Kenneth C. and {Suyu}, Sherry H. and {Auger}, Matthew W. and {Bonvin}, Vivien and {Courbin}, Frederic and {Fassnacht}, Christopher D. and {Halkola}, Aleksi and {Rusu}, Cristian E. and {Sluse}, Dominique and {Sonnenfeld}, Alessandro and {Treu}, Tommaso and {Collett}, Thomas E. and {Hilbert}, Stefan and {Koopmans}, Leon V.~E. and {Marshall}, Philip J. and {Rumbaugh}, Nicholas},
        title = "{H0LiCOW - IV. Lens mass model of HE 0435-1223 and blind measurement of its time-delay distance for cosmology}",
      journal = {\mnras},
         year = 2017,
        month = mar,
       volume = {465},
       number = {4},
        pages = {4895-4913},
          doi = {10.1093/mnras/stw3077},
archivePrefix = {arXiv},
       eprint = {1607.01403},
 primaryClass = {astro-ph.CO},
       adsurl = {https://ui.adsabs.harvard.edu/abs/2017MNRAS.465.4895W}
}

@article{AIC,
author = {Akaike, H.},
year = {1974},
month = {01},
pages = {716-723},
title = {A New Look at the Statistical Model Identification},
volume = {19},
journal = {IEEE Trans Autom Contr}
}

@article{BIC,
author = {Schwarz, Gideon},
year = {1978},
month = {03},
pages = {},
title = {Estimating the Dimension of a Model},
volume = {6},
journal = {The Annals of Statistics},
doi = {10.1214/aos/1176344136}
}

@ARTICLE{BAO,
       author = {{Eisenstein}, Daniel J. and {Zehavi}, Idit and {Hogg}, David W. and {Scoccimarro}, Roman and {Blanton}, Michael R. and {Nichol}, Robert C. and {Scranton}, Ryan and {Seo}, Hee-Jong and {Tegmark}, Max and {Zheng}, Zheng and {Anderson}, Scott F. and {Annis}, Jim and {Bahcall}, Neta and {Brinkmann}, Jon and {Burles}, Scott and {Castander}, Francisco J. and {Connolly}, Andrew and {Csabai}, Istvan and {Doi}, Mamoru and {Fukugita}, Masataka and {Frieman}, Joshua A. and {Glazebrook}, Karl and {Gunn}, James E. and {Hendry}, John S. and {Hennessy}, Gregory and {Ivezi{\'c}}, Zeljko and {Kent}, Stephen and {Knapp}, Gillian R. and {Lin}, Huan and {Loh}, Yeong-Shang and {Lupton}, Robert H. and {Margon}, Bruce and {McKay}, Timothy A. and {Meiksin}, Avery and {Munn}, Jeffery A. and {Pope}, Adrian and {Richmond}, Michael W. and {Schlegel}, David and {Schneider}, Donald P. and {Shimasaku}, Kazuhiro and {Stoughton}, Christopher and {Strauss}, Michael A. and {SubbaRao}, Mark and {Szalay}, Alexander S. and {Szapudi}, Istv{\'a}n and {Tucker}, Douglas L. and {Yanny}, Brian and {York}, Donald G.},
        title = "{Detection of the Baryon Acoustic Peak in the Large-Scale Correlation Function of SDSS Luminous Red Galaxies}",
      journal = {\apj},
         year = 2005,
        month = nov,
       volume = {633},
       number = {2},
        pages = {560-574},
          doi = {10.1086/466512},
archivePrefix = {arXiv},
       eprint = {astro-ph/0501171},
 primaryClass = {astro-ph},
       adsurl = {https://ui.adsabs.harvard.edu/abs/2005ApJ...633..560E}
}

@ARTICLE{2003ApJ...593..622J,
       author = {{Jimenez}, Raul and {Verde}, Licia and {Treu}, Tommaso and {Stern}, Daniel},
        title = "{Constraints on the Equation of State of Dark Energy and the Hubble Constant from Stellar Ages and the Cosmic Microwave Background}",
      journal = {\apj},
         year = 2003,
        month = aug,
       volume = {593},
       number = {2},
        pages = {622-629},
          doi = {10.1086/376595},
archivePrefix = {arXiv},
       eprint = {astro-ph/0302560},
 primaryClass = {astro-ph},
       adsurl = {https://ui.adsabs.harvard.edu/abs/2003ApJ...593..622J}
}

@ARTICLE{2005PhRvD..71l3001S,
       author = {{Simon}, Joan and {Verde}, Licia and {Jimenez}, Raul},
        title = "{Constraints on the redshift dependence of the dark energy potential}",
      journal = {\prd},
         year = 2005,
        month = jun,
       volume = {71},
       number = {12},
          eid = {123001},
        pages = {123001},
          doi = {10.1103/PhysRevD.71.123001},
archivePrefix = {arXiv},
       eprint = {astro-ph/0412269},
 primaryClass = {astro-ph},
       adsurl = {https://ui.adsabs.harvard.edu/abs/2005PhRvD..71l3001S}
}

@ARTICLE{2012JCAP...08..006M,
       author = {{Moresco}, M. and {Cimatti}, A. and {Jimenez}, R. and {Pozzetti}, L. and {Zamorani}, G. and {Bolzonella}, M. and {Dunlop}, J. and {Lamareille}, F. and {Mignoli}, M. and {Pearce}, H. and {Rosati}, P. and {Stern}, D. and {Verde}, L. and {Zucca}, E. and {Carollo}, C.~M. and {Contini}, T. and {Kneib}, J.-P. and {Le F{\`e}vre}, O. and {Lilly}, S.~J. and {Mainieri}, V. and {Renzini}, A. and {Scodeggio}, M. and {Balestra}, I. and {Gobat}, R. and {McLure}, R. and {Bardelli}, S. and {Bongiorno}, A. and {Caputi}, K. and {Cucciati}, O. and {de la Torre}, S. and {de Ravel}, L. and {Franzetti}, P. and {Garilli}, B. and {Iovino}, A. and {Kampczyk}, P. and {Knobel}, C. and {Kova{\v{c}}}, K. and {Le Borgne}, J.-F. and {Le Brun}, V. and {Maier}, C. and {Pell{\'o}}, R. and {Peng}, Y. and {Perez-Montero}, E. and {Presotto}, V. and {Silverman}, J.~D. and {Tanaka}, M. and {Tasca}, L.~A.~M. and {Tresse}, L. and {Vergani}, D. and {Almaini}, O. and {Barnes}, L. and {Bordoloi}, R. and {Bradshaw}, E. and {Cappi}, A. and {Chuter}, R. and {Cirasuolo}, M. and {Coppa}, G. and {Diener}, C. and {Foucaud}, S. and {Hartley}, W. and {Kamionkowski}, M. and {Koekemoer}, A.~M. and {L{\'o}pez-Sanjuan}, C. and {McCracken}, H.~J. and {Nair}, P. and {Oesch}, P. and {Stanford}, A. and {Welikala}, N.},
        title = "{Improved constraints on the expansion rate of the Universe up to z \raisebox{-0.5ex}\textasciitilde 1.1 from the spectroscopic evolution of cosmic chronometers}",
      journal = {\jcap},
         year = 2012,
        month = aug,
       volume = {2012},
       number = {8},
          eid = {006},
        pages = {006},
          doi = {10.1088/1475-7516/2012/08/006},
archivePrefix = {arXiv},
       eprint = {1201.3609},
 primaryClass = {astro-ph.CO},
       adsurl = {https://ui.adsabs.harvard.edu/abs/2012JCAP...08..006M}
}

@ARTICLE{1994ApJS...95..107W,
       author = {{Worthey}, Guy},
        title = "{Comprehensive Stellar Population Models and the Disentanglement of Age and Metallicity Effects}",
      journal = {\apjs},
         year = 1994,
        month = nov,
       volume = {95},
        pages = {107},
          doi = {10.1086/192096},
       adsurl = {https://ui.adsabs.harvard.edu/abs/1994ApJS...95..107W}
}

@ARTICLE{2011MNRAS.412.2183T,
       author = {{Thomas}, Daniel and {Maraston}, Claudia and {Johansson}, Jonas},
        title = "{Flux-calibrated stellar population models of Lick absorption-line indices with variable element abundance ratios}",
      journal = {\mnras},
         year = 2011,
        month = apr,
       volume = {412},
       number = {4},
        pages = {2183-2198},
          doi = {10.1111/j.1365-2966.2010.18049.x},
archivePrefix = {arXiv},
       eprint = {1010.4569},
 primaryClass = {astro-ph.CO},
       adsurl = {https://ui.adsabs.harvard.edu/abs/2011MNRAS.412.2183T}
}

@ARTICLE{2022LRR....25....6M,
       author = {{Moresco}, Michele and {Amati}, Lorenzo and {Amendola}, Luca and {Birrer}, Simon and {Blakeslee}, John P. and {Cantiello}, Michele and {Cimatti}, Andrea and {Darling}, Jeremy and {Della Valle}, Massimo and {Fishbach}, Maya and {Grillo}, Claudio and {Hamaus}, Nico and {Holz}, Daniel and {Izzo}, Luca and {Jimenez}, Raul and {Lusso}, Elisabeta and {Meneghetti}, Massimo and {Piedipalumbo}, Ester and {Pisani}, Alice and {Pourtsidou}, Alkistis and {Pozzetti}, Lucia and {Quartin}, Miguel and {Risaliti}, Guido and {Rosati}, Piero and {Verde}, Licia},
        title = "{Unveiling the Universe with emerging cosmological probes}",
      journal = {Living Reviews in Relativity},
         year = 2022,
        month = dec,
       volume = {25},
       number = {1},
          eid = {6},
        pages = {6},
          doi = {10.1007/s41114-022-00040-z},
archivePrefix = {arXiv},
       eprint = {2201.07241},
 primaryClass = {astro-ph.CO},
       adsurl = {https://ui.adsabs.harvard.edu/abs/2022LRR....25....6M}
}

@ARTICLE{2022ApJ...938..113S,
       author = {{Scolnic}, Dan and {Brout}, Dillon and {Carr}, Anthony and {Riess}, Adam G. and {Davis}, Tamara M. and {Dwomoh}, Arianna and {Jones}, David O. and {Ali}, Noor and {Charvu}, Pranav and {Chen}, Rebecca and {Peterson}, Erik R. and {Popovic}, Brodie and {Rose}, Benjamin M. and {Wood}, Charlotte M. and {Brown}, Peter J. and {Chambers}, Ken and {Coulter}, David A. and {Dettman}, Kyle G. and {Dimitriadis}, Georgios and {Filippenko}, Alexei V. and {Foley}, Ryan J. and {Jha}, Saurabh W. and {Kilpatrick}, Charles D. and {Kirshner}, Robert P. and {Pan}, Yen-Chen and {Rest}, Armin and {Rojas-Bravo}, Cesar and {Siebert}, Matthew R. and {Stahl}, Benjamin E. and {Zheng}, WeiKang},
        title = "{The Pantheon+ Analysis: The Full Data Set and Light-curve Release}",
      journal = {\apj},
         year = 2022,
        month = oct,
       volume = {938},
       number = {2},
          eid = {113},
        pages = {113},
          doi = {10.3847/1538-4357/ac8b7a},
archivePrefix = {arXiv},
       eprint = {2112.03863},
 primaryClass = {astro-ph.CO},
       adsurl = {https://ui.adsabs.harvard.edu/abs/2022ApJ...938..113S}
}

@ARTICLE{2021ApJ...909...26B,
       author = {{Brout}, Dillon and {Scolnic}, Daniel},
        title = "{It's Dust: Solving the Mysteries of the Intrinsic Scatter and Host-galaxy Dependence of Standardized Type Ia Supernova Brightnesses}",
      journal = {\apj},
         year = 2021,
        month = mar,
       volume = {909},
       number = {1},
          eid = {26},
        pages = {26},
          doi = {10.3847/1538-4357/abd69b},
archivePrefix = {arXiv},
       eprint = {2004.10206},
 primaryClass = {astro-ph.CO},
       adsurl = {https://ui.adsabs.harvard.edu/abs/2021ApJ...909...26B}
}

@ARTICLE{2021ApJ...913...49P,
       author = {{Popovic}, Brodie and {Brout}, Dillon and {Kessler}, Richard and {Scolnic}, Dan and {Lu}, Lisa},
        title = "{Improved Treatment of Host-galaxy Correlations in Cosmological Analyses with Type Ia Supernovae}",
      journal = {\apj},
         year = 2021,
        month = may,
       volume = {913},
       number = {1},
          eid = {49},
        pages = {49},
          doi = {10.3847/1538-4357/abf14f},
archivePrefix = {arXiv},
       eprint = {2102.01776},
 primaryClass = {astro-ph.CO},
       adsurl = {https://ui.adsabs.harvard.edu/abs/2021ApJ...913...49P}
}

@ARTICLE{2023ApJ...945...84P,
       author = {{Popovic}, Brodie and {Brout}, Dillon and {Kessler}, Richard and {Scolnic}, Daniel},
        title = "{The Pantheon+ Analysis: Forward Modeling the Dust and Intrinsic Color Distributions of Type Ia Supernovae, and Quantifying Their Impact on Cosmological Inferences}",
      journal = {\apj},
         year = 2023,
        month = mar,
       volume = {945},
       number = {1},
          eid = {84},
        pages = {84},
          doi = {10.3847/1538-4357/aca273},
archivePrefix = {arXiv},
       eprint = {2112.04456},
 primaryClass = {astro-ph.CO},
       adsurl = {https://ui.adsabs.harvard.edu/abs/2023ApJ...945...84P}
}

@ARTICLE{2019ApJ...874..150B,
       author = {{Brout}, D. and {Scolnic}, D. and {Kessler}, R. and {D'Andrea}, C.~B. and {Davis}, T.~M. and {Gupta}, R.~R. and {Hinton}, S.~R. and {Kim}, A.~G. and {Lasker}, J. and {Lidman}, C. and {Macaulay}, E. and {M{\"o}ller}, A. and {Nichol}, R.~C. and {Sako}, M. and {Smith}, M. and {Sullivan}, M. and {Zhang}, B. and {Andersen}, P. and {Asorey}, J. and {Avelino}, A. and {Bassett}, B.~A. and {Brown}, P. and {Calcino}, J. and {Carollo}, D. and {Challis}, P. and {Childress}, M. and {Clocchiatti}, A. and {Filippenko}, A.~V. and {Foley}, R.~J. and {Galbany}, L. and {Glazebrook}, K. and {Hoormann}, J.~K. and {Kasai}, E. and {Kirshner}, R.~P. and {Kuehn}, K. and {Kuhlmann}, S. and {Lewis}, G.~F. and {Mandel}, K.~S. and {March}, M. and {Miranda}, V. and {Morganson}, E. and {Muthukrishna}, D. and {Nugent}, P. and {Palmese}, A. and {Pan}, Y.-C. and {Sharp}, R. and {Sommer}, N.~E. and {Swann}, E. and {Thomas}, R.~C. and {Tucker}, B.~E. and {Uddin}, S.~A. and {Wester}, W. and {Abbott}, T.~M.~C. and {Allam}, S. and {Annis}, J. and {Avila}, S. and {Bechtol}, K. and {Bernstein}, G.~M. and {Bertin}, E. and {Brooks}, D. and {Burke}, D.~L. and {Carnero Rosell}, A. and {Carrasco Kind}, M. and {Carretero}, J. and {Castander}, F.~J. and {Cunha}, C.~E. and {da Costa}, L.~N. and {Davis}, C. and {De Vicente}, J. and {DePoy}, D.~L. and {Desai}, S. and {Diehl}, H.~T. and {Doel}, P. and {Drlica-Wagner}, A. and {Eifler}, T.~F. and {Estrada}, J. and {Fernandez}, E. and {Flaugher}, B. and {Fosalba}, P. and {Frieman}, J. and {Garc{\'\i}a-Bellido}, J. and {Gruen}, D. and {Gruendl}, R.~A. and {Gutierrez}, G. and {Hartley}, W.~G. and {Hollowood}, D.~L. and {Honscheid}, K. and {Hoyle}, B. and {James}, D.~J. and {Jarvis}, M. and {Jeltema}, T. and {Krause}, E. and {Lahav}, O. and {Li}, T.~S. and {Lima}, M. and {Maia}, M.~A.~G. and {Marriner}, J. and {Marshall}, J.~L. and {Martini}, P. and {Menanteau}, F. and {Miller}, C.~J. and {Miquel}, R. and {Ogando}, R.~L.~C. and {Plazas}, A.~A. and {Romer}, A.~K. and {Roodman}, A. and {Rykoff}, E.~S. and {Sanchez}, E. and {Santiago}, B. and {Scarpine}, V. and {Schubnell}, M. and {Serrano}, S. and {Sevilla-Noarbe}, I. and {Smith}, R.~C. and {Soares-Santos}, M. and {Sobreira}, F. and {Suchyta}, E. and {Swanson}, M.~E.~C. and {Tarle}, G. and {Thomas}, D. and {Troxel}, M.~A. and {Tucker}, D.~L. and {Vikram}, V. and {Walker}, A.~R. and {Zhang}, Y. and {DES Collaboration}},
        title = "{First Cosmology Results Using SNe Ia from the Dark Energy Survey: Analysis, Systematic Uncertainties, and Validation}",
      journal = {\apj},
         year = 2019,
        month = apr,
       volume = {874},
       number = {2},
          eid = {150},
        pages = {150},
          doi = {10.3847/1538-4357/ab08a0},
archivePrefix = {arXiv},
       eprint = {1811.02377},
 primaryClass = {astro-ph.CO},
       adsurl = {https://ui.adsabs.harvard.edu/abs/2019ApJ...874..150B}
}

@ARTICLE{2022ApJ...938..112P,
       author = {{Peterson}, Erik R. and {Kenworthy}, W. D'Arcy and {Scolnic}, Daniel and {Riess}, Adam G. and {Brout}, Dillon and {Carr}, Anthony and {Courtois}, H{\'e}l{\`e}ne and {Davis}, Tamara and {Dwomoh}, Arianna and {Jones}, David O. and {Popovic}, Brodie and {Rose}, Benjamin M. and {Said}, Khaled},
        title = "{The Pantheon+ Analysis: Evaluating Peculiar Velocity Corrections in Cosmological Analyses with Nearby Type Ia Supernovae}",
      journal = {\apj},
         year = 2022,
        month = oct,
       volume = {938},
       number = {2},
          eid = {112},
        pages = {112},
          doi = {10.3847/1538-4357/ac4698},
archivePrefix = {arXiv},
       eprint = {2110.03487},
 primaryClass = {astro-ph.CO},
       adsurl = {https://ui.adsabs.harvard.edu/abs/2022ApJ...938..112P}
}

@ARTICLE{2024ApJ...973L..14D,
       author = {{DES Collaboration} and {Abbott}, T.~M.~C. and {Acevedo}, M. and {Aguena}, M. and {Alarcon}, A. and {Allam}, S. and {Alves}, O. and {Amon}, A. and {Andrade-Oliveira}, F. and {Annis}, J. and {Armstrong}, P. and {Asorey}, J. and {Avila}, S. and {Bacon}, D. and {Bassett}, B.~A. and {Bechtol}, K. and {Bernardinelli}, P.~H. and {Bernstein}, G.~M. and {Bertin}, E. and {Blazek}, J. and {Bocquet}, S. and {Brooks}, D. and {Brout}, D. and {Buckley-Geer}, E. and {Burke}, D.~L. and {Camacho}, H. and {Camilleri}, R. and {Campos}, A. and {Carnero Rosell}, A. and {Carollo}, D. and {Carr}, A. and {Carretero}, J. and {Castander}, F.~J. and {Cawthon}, R. and {Chang}, C. and {Chen}, R. and {Choi}, A. and {Conselice}, C. and {Costanzi}, M. and {da Costa}, L.~N. and {Crocce}, M. and {Davis}, T.~M. and {DePoy}, D.~L. and {Desai}, S. and {Diehl}, H.~T. and {Dixon}, M. and {Dodelson}, S. and {Doel}, P. and {Doux}, C. and {Drlica-Wagner}, A. and {Elvin-Poole}, J. and {Everett}, S. and {Ferrero}, I. and {Fert{\'e}}, A. and {Flaugher}, B. and {Foley}, R.~J. and {Fosalba}, P. and {Friedel}, D. and {Frieman}, J. and {Frohmaier}, C. and {Galbany}, L. and {Garc{\'\i}a-Bellido}, J. and {Gatti}, M. and {Gaztanaga}, E. and {Giannini}, G. and {Glazebrook}, K. and {Graur}, O. and {Gruen}, D. and {Gruendl}, R.~A. and {Gutierrez}, G. and {Hartley}, W.~G. and {Herner}, K. and {Hinton}, S.~R. and {Hollowood}, D.~L. and {Honscheid}, K. and {Huterer}, D. and {Jain}, B. and {James}, D.~J. and {Jeffrey}, N. and {Kasai}, E. and {Kelsey}, L. and {Kent}, S. and {Kessler}, R. and {Kim}, A.~G. and {Kirshner}, R.~P. and {Kovacs}, E. and {Kuehn}, K. and {Lahav}, O. and {Lee}, J. and {Lee}, S. and {Lewis}, G.~F. and {Li}, T.~S. and {Lidman}, C. and {Lin}, H. and {Malik}, U. and {Marshall}, J.~L. and {Martini}, P. and {Mena-Fern{\'a}ndez}, J. and {Menanteau}, F. and {Miquel}, R. and {Mohr}, J.~J. and {Mould}, J. and {Muir}, J. and {M{\"o}ller}, A. and {Neilsen}, E. and {Nichol}, R.~C. and {Nugent}, P. and {Ogando}, R.~L.~C. and {Palmese}, A. and {Pan}, Y.-C. and {Paterno}, M. and {Percival}, W.~J. and {Pereira}, M.~E.~S. and {Pieres}, A. and {Malag{\'o}n}, A.~A. Plazas and {Popovic}, B. and {Porredon}, A. and {Prat}, J. and {Qu}, H. and {Raveri}, M. and {Rodr{\'\i}guez-Monroy}, M. and {Romer}, A.~K. and {Roodman}, A. and {Rose}, B. and {Sako}, M. and {Sanchez}, E. and {Sanchez Cid}, D. and {Schubnell}, M. and {Scolnic}, D. and {Sevilla-Noarbe}, I. and {Shah}, P. and {Smith}, J. Allyn. and {Smith}, M. and {Soares-Santos}, M. and {Suchyta}, E. and {Sullivan}, M. and {Suntzeff}, N. and {Swanson}, M.~E.~C. and {S{\'a}nchez}, B.~O. and {Tarle}, G. and {Taylor}, G. and {Thomas}, D. and {To}, C. and {Toy}, M. and {Troxel}, M.~A. and {Tucker}, B.~E. and {Tucker}, D.~L. and {Uddin}, S.~A. and {Vincenzi}, M. and {Walker}, A.~R. and {Weaverdyck}, N. and {Wechsler}, R.~H. and {Weller}, J. and {Wester}, W. and {Wiseman}, P. and {Yamamoto}, M. and {Yuan}, F. and {Zhang}, B. and {Zhang}, Y.},
        title = "{The Dark Energy Survey: Cosmology Results with {\ensuremath{\sim}}1500 New High-redshift Type Ia Supernovae Using the Full 5 yr Data Set}",
      journal = {\apjl},
         year = 2024,
        month = sep,
       volume = {973},
       number = {1},
          eid = {L14},
        pages = {L14},
          doi = {10.3847/2041-8213/ad6f9f},
archivePrefix = {arXiv},
       eprint = {2401.02929},
 primaryClass = {astro-ph.CO},
       adsurl = {https://ui.adsabs.harvard.edu/abs/2024ApJ...973L..14D}
}

@article{Kessler_2017,
doi = {10.3847/1538-4357/836/1/56},
url = {https://doi.org/10.3847/1538-4357/836/1/56},
year = {2017},
month = {feb},
publisher = {The American Astronomical Society},
volume = {836},
number = {1},
pages = {56},
author = {Kessler, R. and Scolnic, D.},
title = {Correcting Type Ia Supernova Distances for Selection Biases and Contamination in Photometrically Identified Samples},
journal = {The Astrophysical Journal}}

@ARTICLE{2020MNRAS.491.4277M,
       author = {{M{\"o}ller}, A. and {de Boissi{\`e}re}, T.},
        title = "{SuperNNova: an open-source framework for Bayesian, neural network-based supernova classification}",
      journal = {\mnras},
         year = 2020,
        month = jan,
       volume = {491},
       number = {3},
        pages = {4277-4293},
          doi = {10.1093/mnras/stz3312},
archivePrefix = {arXiv},
       eprint = {1901.06384},
 primaryClass = {astro-ph.IM},
       adsurl = {https://ui.adsabs.harvard.edu/abs/2020MNRAS.491.4277M}
}

@INCOLLECTION{2010dken.book..246B,
       author = {{Bassett}, Bruce and {Hlozek}, Renee},
        title = "{Baryon acoustic oscillations}",
    booktitle = {Dark Energy},
         year = 2010,
       editor = {{Ruiz-Lapuente}, Pilar},
        pages = {246},
          doi = {10.48550/arXiv.0910.5224},
       adsurl = {https://ui.adsabs.harvard.edu/abs/2010dken.book..246B}
}

@ARTICLE{1998ApJ...496..605E,
       author = {{Eisenstein}, Daniel J. and {Hu}, Wayne},
        title = "{Baryonic Features in the Matter Transfer Function}",
      journal = {\apj},
         year = 1998,
        month = mar,
       volume = {496},
       number = {2},
        pages = {605-614},
          doi = {10.1086/305424},
archivePrefix = {arXiv},
       eprint = {astro-ph/9709112},
 primaryClass = {astro-ph},
       adsurl = {https://ui.adsabs.harvard.edu/abs/1998ApJ...496..605E}
}

@ARTICLE{1972CoASP...4..173S,
       author = {{Sunyaev}, R.~A. and {Zeldovich}, Ya. B.},
        title = "{The Observations of Relic Radiation as a Test of the Nature of X-Ray Radiation from the Clusters of Galaxies}",
      journal = {Comments on Astrophysics and Space Physics},
         year = 1972,
        month = nov,
       volume = {4},
        pages = {173},
       adsurl = {https://ui.adsabs.harvard.edu/abs/1972CoASP...4..173S}
}

@ARTICLE{1970ApJ...162..815P,
       author = {{Peebles}, P.~J.~E. and {Yu}, J.~T.},
        title = "{Primeval Adiabatic Perturbation in an Expanding Universe}",
      journal = {\apj},
         year = 1970,
        month = dec,
       volume = {162},
        pages = {815},
          doi = {10.1086/150713},
       adsurl = {https://ui.adsabs.harvard.edu/abs/1970ApJ...162..815P}
}

@ARTICLE{2012MNRAS.426.1280S,
       author = {{Sutherland}, Will},
        title = "{On measuring the absolute scale of baryon acoustic oscillations}",
      journal = {\mnras},
         year = 2012,
        month = oct,
       volume = {426},
       number = {2},
        pages = {1280-1290},
          doi = {10.1111/j.1365-2966.2012.21666.x},
archivePrefix = {arXiv},
       eprint = {1205.0715},
 primaryClass = {astro-ph.CO},
       adsurl = {https://ui.adsabs.harvard.edu/abs/2012MNRAS.426.1280S}
}

@ARTICLE{2007ApJ...664..675E,
       author = {{Eisenstein}, Daniel J. and {Seo}, Hee-Jong and {Sirko}, Edwin and {Spergel}, David N.},
        title = "{Improving Cosmological Distance Measurements by Reconstruction of the Baryon Acoustic Peak}",
      journal = {\apj},
         year = 2007,
        month = aug,
       volume = {664},
       number = {2},
        pages = {675-679},
          doi = {10.1086/518712},
archivePrefix = {arXiv},
       eprint = {astro-ph/0604362},
 primaryClass = {astro-ph},
       adsurl = {https://ui.adsabs.harvard.edu/abs/2007ApJ...664..675E}
}

@ARTICLE{2025PhRvD.112h3511L,
       author = {{Lodha}, K. and {Calderon}, R. and {Matthewson}, W.~L. and {Shafieloo}, A. and {Ishak}, M. and {Pan}, J. and {Garcia-Quintero}, C. and {Huterer}, D. and {Valogiannis}, G. and {Ure{\~n}a-L{\'o}pez}, L.~A. and {Kamble}, N.~V. and {Parkinson}, D. and {Kim}, A.~G. and {Zhao}, G.~B. and {Cervantes-Cota}, J.~L. and {Rohlf}, J. and {Lozano-Rodr{\'\i}guez}, F. and {Rom{\'a}n-Herrera}, J.~O. and {Abdul-Karim}, M. and {Aguilar}, J. and {Ahlen}, S. and {Alves}, O. and {Andrade}, U. and {Armengaud}, E. and {Aviles}, A. and {Behera}, J. and {BenZvi}, S. and {Bianchi}, D. and {Brodzeller}, A. and {Brooks}, D. and {Burtin}, E. and {Canning}, R. and {Rosell}, A. Carnero and {Casas}, L. and {Castander}, F.~J. and {Charles}, M. and {Chaussidon}, E. and {Chaves-Montero}, J. and {Chebat}, D. and {Claybaugh}, T. and {Cole}, S. and {Cuceu}, A. and {Dawson}, K.~S. and {de la Macorra}, A. and {de Mattia}, A. and {Deiosso}, N. and {Demina}, R. and {Dey}, Arjun and {Dey}, Biprateep and {Ding}, Z. and {Doel}, P. and {Eisenstein}, D.~J. and {Elbers}, W. and {Ferraro}, S. and {Font-Ribera}, A. and {Forero-Romero}, J.~E. and {Garrison}, Lehman H. and {Gazta{\~n}aga}, E. and {Gil-Mar{\'\i}n}, H. and {Gontcho}, S. Gontcho A. and {Gonzalez-Morales}, A.~X. and {Gutierrez}, G. and {Guy}, J. and {Hahn}, C. and {Herbold}, M. and {Herrera-Alcantar}, H.~K. and {Honscheid}, K. and {Howlett}, C. and {Juneau}, S. and {Kehoe}, R. and {Kirkby}, D. and {Kisner}, T. and {Kremin}, A. and {Lahav}, O. and {Lamman}, C. and {Landriau}, M. and {Le Guillou}, L. and {Leauthaud}, A. and {Levi}, M.~E. and {Li}, Q. and {Magneville}, C. and {Manera}, M. and {Martini}, P. and {Meisner}, A. and {Mena-Fern{\'a}ndez}, J. and {Miquel}, R. and {Moustakas}, J. and {Santos}, D. Mu{\~n}oz and {Mu{\~n}oz-Guti{\'e}rrez}, A. and {Myers}, A.~D. and {Nadathur}, S. and {Niz}, G. and {Noriega}, H.~E. and {Paillas}, E. and {Palanque-Delabrouille}, N. and {Percival}, W.~J. and {Pieri}, Matthew M. and {Poppett}, C. and {Prada}, F. and {P{\'e}rez-Fern{\'a}ndez}, A. and {P{\'e}rez-R{\`a}fols}, I. and {Ram{\'\i}rez-P{\'e}rez}, C. and {Rashkovetskyi}, M. and {Ravoux}, C. and {Ross}, A.~J. and {Rossi}, G. and {Ruhlmann-Kleider}, V. and {Samushia}, L. and {Sanchez}, E. and {Schlegel}, D. and {Schubnell}, M. and {Seo}, H. and {Sinigaglia}, F. and {Sprayberry}, D. and {Tan}, T. and {Tarl{\'e}}, G. and {Taylor}, P. and {Turner}, W. and {Vargas-Maga{\~n}a}, M. and {Walther}, M. and {Weaver}, B.~A. and {Wolfson}, M. and {Y{\`e}che}, C. and {Zarrouk}, P. and {Zhou}, R. and {Zou}, H. and {DESI Collaboration}},
        title = "{Extended dark energy analysis using DESI DR2 BAO measurements}",
      journal = {\prd},
         year = 2025,
        month = oct,
       volume = {112},
       number = {8},
          eid = {083511},
        pages = {083511},
          doi = {10.1103/w4c6-1r5j},
archivePrefix = {arXiv},
       eprint = {2503.14743},
 primaryClass = {astro-ph.CO},
       adsurl = {https://ui.adsabs.harvard.edu/abs/2025PhRvD.112h3511L}
}

@ARTICLE{2025PhRvD.112h3515A,
       author = {{Abdul Karim}, M. and {Aguilar}, J. and {Ahlen}, S. and {Alam}, S. and {Allen}, L. and {Prieto}, C. Allende and {Alves}, O. and {Anand}, A. and {Andrade}, U. and {Armengaud}, E. and {Aviles}, A. and {Bailey}, S. and {Baltay}, C. and {Bansal}, P. and {Bault}, A. and {Behera}, J. and {BenZvi}, S. and {Bianchi}, D. and {Blake}, C. and {Brieden}, S. and {Brodzeller}, A. and {Brooks}, D. and {Buckley-Geer}, E. and {Burtin}, E. and {Calderon}, R. and {Canning}, R. and {Rosell}, A. Carnero and {Carrilho}, P. and {Casas}, L. and {Castander}, F.~J. and {Charles}, M. and {Chaussidon}, E. and {Chaves-Montero}, J. and {Chebat}, D. and {Chen}, X. and {Claybaugh}, T. and {Cole}, S. and {Cooper}, A.~P. and {Cuceu}, A. and {Dawson}, K.~S. and {de la Macorra}, A. and {de Mattia}, A. and {Deiosso}, N. and {Della Costa}, J. and {Demina}, R. and {Dey}, A. and {Dey}, B. and {Ding}, Z. and {Doel}, P. and {Edelstein}, J. and {Eisenstein}, D.~J. and {Elbers}, W. and {Fagrelius}, P. and {Fanning}, K. and {Fern{\'a}ndez-Garc{\'\i}a}, E. and {Ferraro}, S. and {Font-Ribera}, A. and {Forero-Romero}, J.~E. and {Frenk}, C.~S. and {Garcia-Quintero}, C. and {Garrison}, L.~H. and {Gazta{\~n}aga}, E. and {Gil-Mar{\'\i}n}, H. and {Gontcho A Gontcho}, S. and {Gonzalez}, D. and {Gonzalez-Morales}, A.~X. and {Gordon}, C. and {Green}, D. and {Gutierrez}, G. and {Guy}, J. and {Hadzhiyska}, B. and {Hahn}, C. and {He}, S. and {Herbold}, M. and {Herrera-Alcantar}, H.~K. and {Ho}, M.-F. and {Honscheid}, K. and {Howlett}, C. and {Huterer}, D. and {Ishak}, M. and {Juneau}, S. and {Kamble}, N.~V. and {Kara{\c{c}}ayl{\i}}, N.~G. and {Kehoe}, R. and {Kent}, S. and {Kim}, A.~G. and {Kirkby}, D. and {Kisner}, T. and {Koposov}, S.~E. and {Kremin}, A. and {Krolewski}, A. and {Lahav}, O. and {Lamman}, C. and {Landriau}, M. and {Lang}, D. and {Lasker}, J. and {Le Goff}, J.~M. and {Le Guillou}, L. and {Leauthaud}, A. and {Levi}, M.~E. and {Li}, Q. and {Li}, T.~S. and {Lodha}, K. and {Lokken}, M. and {Lozano-Rodr{\'\i}guez}, F. and {Magneville}, C. and {Manera}, M. and {Martini}, P. and {Matthewson}, W.~L. and {Meisner}, A. and {Mena-Fern{\'a}ndez}, J. and {Menegas}, A. and {Mergulh{\~a}o}, T. and {Miquel}, R. and {Moustakas}, J. and {Mu{\~n}oz-Guti{\'e}rrez}, A. and {Mu{\~n}oz-Santos}, D. and {Myers}, A.~D. and {Nadathur}, S. and {Naidoo}, K. and {Napolitano}, L. and {Newman}, J.~A. and {Niz}, G. and {Noriega}, H.~E. and {Paillas}, E. and {Palanque-Delabrouille}, N. and {Pan}, J. and {Peacock}, J.~A. and {Pellejero Ibanez}, M. and {Percival}, W.~J. and {P{\'e}rez-Fern{\'a}ndez}, A. and {P{\'e}rez-R{\`a}fols}, I. and {Pieri}, M.~M. and {Poppett}, C. and {Prada}, F. and {Rabinowitz}, D. and {Raichoor}, A. and {Ram{\'\i}rez-P{\'e}rez}, C. and {Rashkovetskyi}, M. and {Ravoux}, C. and {Rich}, J. and {Rocher}, A. and {Rockosi}, C. and {Rohlf}, J. and {Rom{\'a}n-Herrera}, J.~O. and {Ross}, A.~J. and {Rossi}, G. and {Ruggeri}, R. and {Ruhlmann-Kleider}, V. and {Samushia}, L. and {Sanchez}, E. and {Sanders}, N. and {Schlegel}, D. and {Schubnell}, M. and {Seo}, H. and {Shafieloo}, A. and {Sharples}, R. and {Silber}, J. and {Sinigaglia}, F. and {Sprayberry}, D. and {Tan}, T. and {Tarl{\'e}}, G. and {Taylor}, P. and {Turner}, W. and {Ure{\~n}a-L{\'o}pez}, L.~A. and {Vaisakh}, R. and {Valdes}, F. and {Valogiannis}, G. and {Vargas-Maga{\~n}a}, M. and {Verde}, L. and {Walther}, M. and {Weaver}, B.~A. and {Weinberg}, D.~H. and {White}, M. and {Wolfson}, M. and {Y{\`e}che}, C. and {Yu}, J. and {Zaborowski}, E.~A. and {Zarrouk}, P. and {Zhai}, Z. and {Zhang}, H. and {Zhao}, C. and {Zhao}, G.~B. and {Zhou}, R. and {Zou}, H. and {DESI Collaboration}},
        title = "{DESI DR2 results. II. Measurements of baryon acoustic oscillations and cosmological constraints}",
      journal = {\prd},
         year = 2025,
        month = oct,
       volume = {112},
       number = {8},
          eid = {083515},
        pages = {083515},
          doi = {10.1103/tr6y-kpc6},
archivePrefix = {arXiv},
       eprint = {2503.14738},
 primaryClass = {astro-ph.CO},
       adsurl = {https://ui.adsabs.harvard.edu/abs/2025PhRvD.112h3515A}
}

@ARTICLE{1990CQGra...7.1319P,
       author = {{Perlick}, V.},
        title = "{On Fermat's principle in general relativity. I. The general case}",
      journal = {Classical and Quantum Gravity},
         year = 1990,
        month = aug,
       volume = {7},
       number = {8},
        pages = {1319-1331},
          doi = {10.1088/0264-9381/7/8/011},
       adsurl = {https://ui.adsabs.harvard.edu/abs/1990CQGra...7.1319P}
}

@ARTICLE{1990CQGra...7.1849P,
       author = {{Perlick}, V.},
        title = "{On Fermat's principle in general relativity. II. The conformally stationary case}",
      journal = {Classical and Quantum Gravity},
         year = 1990,
        month = oct,
       volume = {7},
       number = {10},
        pages = {1849-1867},
          doi = {10.1088/0264-9381/7/10/016},
       adsurl = {https://ui.adsabs.harvard.edu/abs/1990CQGra...7.1849P}
}

@ARTICLE{2016A&ARv..24...11T,
       author = {{Treu}, Tommaso and {Marshall}, Philip J.},
        title = "{Time delay cosmography}",
      journal = {\aapr},
         year = 2016,
        month = dec,
       volume = {24},
       number = {1},
          eid = {11},
        pages = {11},
          doi = {10.1007/s00159-016-0096-8},
archivePrefix = {arXiv},
       eprint = {1605.05333},
 primaryClass = {astro-ph.CO},
       adsurl = {https://ui.adsabs.harvard.edu/abs/2016A&ARv..24...11T}
}

@ARTICLE{1994ApJ...436..509S,
       author = {{Seljak}, Uros},
        title = "{Large-Scale Structure Effects on the Gravitational Lens Image Positions and Time Delay}",
      journal = {\apj},
         year = 1994,
        month = dec,
       volume = {436},
        pages = {509},
          doi = {10.1086/174924},
archivePrefix = {arXiv},
       eprint = {astro-ph/9405002},
 primaryClass = {astro-ph},
       adsurl = {https://ui.adsabs.harvard.edu/abs/1994ApJ...436..509S}
}

@ARTICLE{2003ApJ...584..664K,
       author = {{Keeton}, Charles R.},
        title = "{Analytic Cross Sections for Substructure Lensing}",
      journal = {\apj},
         year = 2003,
        month = feb,
       volume = {584},
       number = {2},
        pages = {664-674},
          doi = {10.1086/345717},
archivePrefix = {arXiv},
       eprint = {astro-ph/0209040},
 primaryClass = {astro-ph},
       adsurl = {https://ui.adsabs.harvard.edu/abs/2003ApJ...584..664K}
}

@ARTICLE{2014MNRAS.443.3631M,
       author = {{McCully}, Curtis and {Keeton}, Charles R. and {Wong}, Kenneth C. and {Zabludoff}, Ann I.},
        title = "{A new hybrid framework to efficiently model lines of sight to gravitational lenses}",
      journal = {\mnras},
         year = 2014,
        month = oct,
       volume = {443},
       number = {4},
        pages = {3631-3642},
          doi = {10.1093/mnras/stu1316},
archivePrefix = {arXiv},
       eprint = {1401.0197},
 primaryClass = {astro-ph.CO},
       adsurl = {https://ui.adsabs.harvard.edu/abs/2014MNRAS.443.3631M}
}

@ARTICLE{2019Sci...365.1134J,
       author = {{Jee}, Inh and {Suyu}, Sherry H. and {Komatsu}, Eiichiro and {Fassnacht}, Christopher D. and {Hilbert}, Stefan and {Koopmans}, L{\'e}on V.~E.},
        title = "{A measurement of the Hubble constant from angular diameter distances to two gravitational lenses}",
      journal = {Science},
         year = 2019,
        month = sep,
       volume = {365},
       number = {6458},
        pages = {1134-1138},
          doi = {10.1126/science.aat7371},
archivePrefix = {arXiv},
       eprint = {1909.06712},
 primaryClass = {astro-ph.CO},
       adsurl = {https://ui.adsabs.harvard.edu/abs/2019Sci...365.1134J}
}

@ARTICLE{2013ApJ...766...70S,
       author = {{Suyu}, S.~H. and {Auger}, M.~W. and {Hilbert}, S. and {Marshall}, P.~J. and {Tewes}, M. and {Treu}, T. and {Fassnacht}, C.~D. and {Koopmans}, L.~V.~E. and {Sluse}, D. and {Blandford}, R.~D. and {Courbin}, F. and {Meylan}, G.},
        title = "{Two Accurate Time-delay Distances from Strong Lensing: Implications for Cosmology}",
      journal = {\apj},
         year = 2013,
        month = apr,
       volume = {766},
       number = {2},
          eid = {70},
        pages = {70},
          doi = {10.1088/0004-637X/766/2/70},
archivePrefix = {arXiv},
       eprint = {1208.6010},
 primaryClass = {astro-ph.CO},
       adsurl = {https://ui.adsabs.harvard.edu/abs/2013ApJ...766...70S}
}

@ARTICLE{2010ApJ...711..201S,
       author = {{Suyu}, S.~H. and {Marshall}, P.~J. and {Auger}, M.~W. and {Hilbert}, S. and {Blandford}, R.~D. and {Koopmans}, L.~V.~E. and {Fassnacht}, C.~D. and {Treu}, T.},
        title = "{Dissecting the Gravitational lens B1608+656. II. Precision Measurements of the Hubble Constant, Spatial Curvature, and the Dark Energy Equation of State}",
      journal = {\apj},
         year = 2010,
        month = mar,
       volume = {711},
       number = {1},
        pages = {201-221},
          doi = {10.1088/0004-637X/711/1/201},
archivePrefix = {arXiv},
       eprint = {0910.2773},
 primaryClass = {astro-ph.CO},
       adsurl = {https://ui.adsabs.harvard.edu/abs/2010ApJ...711..201S}
}

@ARTICLE{2014ApJ...788L..35S,
       author = {{Suyu}, S.~H. and {Treu}, T. and {Hilbert}, S. and {Sonnenfeld}, A. and {Auger}, M.~W. and {Blandford}, R.~D. and {Collett}, T. and {Courbin}, F. and {Fassnacht}, C.~D. and {Koopmans}, L.~V.~E. and {Marshall}, P.~J. and {Meylan}, G. and {Spiniello}, C. and {Tewes}, M.},
        title = "{Cosmology from Gravitational Lens Time Delays and Planck Data}",
      journal = {\apjl},
         year = 2014,
        month = jun,
       volume = {788},
       number = {2},
          eid = {L35},
        pages = {L35},
          doi = {10.1088/2041-8205/788/2/L35},
archivePrefix = {arXiv},
       eprint = {1306.4732},
 primaryClass = {astro-ph.CO},
       adsurl = {https://ui.adsabs.harvard.edu/abs/2014ApJ...788L..35S}
}

@ARTICLE{2020MNRAS.498.1440R,
       author = {{Rusu}, Cristian E. and {Wong}, Kenneth C. and {Bonvin}, Vivien and {Sluse}, Dominique and {Suyu}, Sherry H. and {Fassnacht}, Christopher D. and {Chan}, James H.~H. and {Hilbert}, Stefan and {Auger}, Matthew W. and {Sonnenfeld}, Alessandro and {Birrer}, Simon and {Courbin}, Frederic and {Treu}, Tommaso and {Chen}, Geoff C. -F. and {Halkola}, Aleksi and {Koopmans}, L{\'e}on V.~E. and {Marshall}, Philip J. and {Shajib}, Anowar J.},
        title = "{H0LiCOW XII. Lens mass model of WFI2033-4723 and blind measurement of its time-delay distance and H$_{0}$}",
      journal = {\mnras},
         year = 2020,
        month = oct,
       volume = {498},
       number = {1},
        pages = {1440-1468},
          doi = {10.1093/mnras/stz3451},
archivePrefix = {arXiv},
       eprint = {1905.09338},
 primaryClass = {astro-ph.CO},
       adsurl = {https://ui.adsabs.harvard.edu/abs/2020MNRAS.498.1440R}
}

@ARTICLE{2019MNRAS.490.1743C,
       author = {{Chen}, Geoff C. -F. and {Fassnacht}, Christopher D. and {Suyu}, Sherry H. and {Rusu}, Cristian E. and {Chan}, James H.~H. and {Wong}, Kenneth C. and {Auger}, Matthew W. and {Hilbert}, Stefan and {Bonvin}, Vivien and {Birrer}, Simon and {Millon}, Martin and {Koopmans}, L{\'e}on V.~E. and {Lagattuta}, David J. and {McKean}, John P. and {Vegetti}, Simona and {Courbin}, Frederic and {Ding}, Xuheng and {Halkola}, Aleksi and {Jee}, Inh and {Shajib}, Anowar J. and {Sluse}, Dominique and {Sonnenfeld}, Alessandro and {Treu}, Tommaso},
        title = "{A SHARP view of H0LiCOW: H$_{0}$ from three time-delay gravitational lens systems with adaptive optics imaging}",
      journal = {\mnras},
         year = 2019,
        month = dec,
       volume = {490},
       number = {2},
        pages = {1743-1773},
          doi = {10.1093/mnras/stz2547},
archivePrefix = {arXiv},
       eprint = {1907.02533},
 primaryClass = {astro-ph.CO},
       adsurl = {https://ui.adsabs.harvard.edu/abs/2019MNRAS.490.1743C}
}

@ARTICLE{2019MNRAS.484.4726B,
       author = {{Birrer}, S. and {Treu}, T. and {Rusu}, C.~E. and {Bonvin}, V. and {Fassnacht}, C.~D. and {Chan}, J.~H.~H. and {Agnello}, A. and {Shajib}, A.~J. and {Chen}, G.~C. -F. and {Auger}, M. and {Courbin}, F. and {Hilbert}, S. and {Sluse}, D. and {Suyu}, S.~H. and {Wong}, K.~C. and {Marshall}, P. and {Lemaux}, B.~C. and {Meylan}, G.},
        title = "{H0LiCOW - IX. Cosmographic analysis of the doubly imaged quasar SDSS 1206+4332 and a new measurement of the Hubble constant}",
      journal = {\mnras},
         year = 2019,
        month = apr,
       volume = {484},
       number = {4},
        pages = {4726-4753},
          doi = {10.1093/mnras/stz200},
archivePrefix = {arXiv},
       eprint = {1809.01274},
 primaryClass = {astro-ph.CO},
       adsurl = {https://ui.adsabs.harvard.edu/abs/2019MNRAS.484.4726B}
}

@ARTICLE{2020ApJ...898...82M,
       author = {{Moresco}, Michele and {Jimenez}, Raul and {Verde}, Licia and {Cimatti}, Andrea and {Pozzetti}, Lucia},
        title = "{Setting the Stage for Cosmic Chronometers. II. Impact of Stellar Population Synthesis Models Systematics and Full Covariance Matrix}",
      journal = {\apj},
         year = 2020,
        month = jul,
       volume = {898},
       number = {1},
          eid = {82},
        pages = {82},
          doi = {10.3847/1538-4357/ab9eb0},
archivePrefix = {arXiv},
       eprint = {2003.07362},
 primaryClass = {astro-ph.GA},
       adsurl = {https://ui.adsabs.harvard.edu/abs/2020ApJ...898...82M}
}

@ARTICLE{1999ChJPh..37..113N,
       author = {{Nester}, James M. and {Yo}, Hwei-Jang},
        title = "{Symmetric Teleparallel General Relativity}",
      journal = {Chinese Journal of Physics},
         year = 1999,
        month = apr,
       volume = {37},
       number = {2},
        pages = {113},
          doi = {10.48550/arXiv.gr-qc/9809049},
archivePrefix = {arXiv},
       eprint = {gr-qc/9809049},
 primaryClass = {gr-qc},
       adsurl = {https://ui.adsabs.harvard.edu/abs/1999ChJPh..37..113N}
}

@article{BeltranJimenez:2017tkd,
    author = "Beltr{\'a}n Jim{\'e}nez, Jose and Heisenberg, Lavinia and Koivisto, Tomi",
    title = "{Coincident General Relativity}",
    eprint = "1710.03116",
    archivePrefix = "arXiv",
    primaryClass = "gr-qc",
    reportNumber = "NORDITA-2017-100, IFT-UAM/CSIC-17-093, ITS-ETH-2017-10",
    doi = "10.1103/PhysRevD.98.044048",
    journal = "Phys. Rev. D",
    volume = "98",
    number = "4",
    pages = "044048",
    year = "2018"
}

@ARTICLE{2020PhRvD.101j3507J,
       author = {{Jim{\'e}nez}, Jose Beltr{\'a}n and {Heisenberg}, Lavinia and {Koivisto}, Tomi and {Pekar}, Simon},
        title = "{Cosmology in f (Q ) geometry}",
      journal = {\prd},
         year = 2020,
        month = may,
       volume = {101},
       number = {10},
          eid = {103507},
        pages = {103507},
          doi = {10.1103/PhysRevD.101.103507},
archivePrefix = {arXiv},
       eprint = {1906.10027},
 primaryClass = {gr-qc},
       adsurl = {https://ui.adsabs.harvard.edu/abs/2020PhRvD.101j3507J}
}

@ARTICLE{2003ApJS..148..233P,
       author = {{Page}, L. and {Nolta}, M.~R. and {Barnes}, C. and {Bennett}, C.~L. and {Halpern}, M. and {Hinshaw}, G. and {Jarosik}, N. and {Kogut}, A. and {Limon}, M. and {Meyer}, S.~S. and {Peiris}, H.~V. and {Spergel}, D.~N. and {Tucker}, G.~S. and {Wollack}, E. and {Wright}, E.~L.},
        title = "{First-Year Wilkinson Microwave Anisotropy Probe (WMAP) Observations: Interpretation of the TT and TE Angular Power Spectrum Peaks}",
      journal = {\apjs},
         year = 2003,
        month = sep,
       volume = {148},
       number = {1},
        pages = {233-241},
          doi = {10.1086/377224},
archivePrefix = {arXiv},
       eprint = {astro-ph/0302220},
 primaryClass = {astro-ph},
       adsurl = {https://ui.adsabs.harvard.edu/abs/2003ApJS..148..233P}
}

@ARTICLE{2007ApJS..170..377S,
       author = {{Spergel}, D.~N. and {Bean}, R. and {Dor{\'e}}, O. and {Nolta}, M.~R. and {Bennett}, C.~L. and {Dunkley}, J. and {Hinshaw}, G. and {Jarosik}, N. and {Komatsu}, E. and {Page}, L. and {Peiris}, H.~V. and {Verde}, L. and {Halpern}, M. and {Hill}, R.~S. and {Kogut}, A. and {Limon}, M. and {Meyer}, S.~S. and {Odegard}, N. and {Tucker}, G.~S. and {Weiland}, J.~L. and {Wollack}, E. and {Wright}, E.~L.},
        title = "{Three-Year Wilkinson Microwave Anisotropy Probe (WMAP) Observations: Implications for Cosmology}",
      journal = {\apjs},
         year = 2007,
        month = jun,
       volume = {170},
       number = {2},
        pages = {377-408},
          doi = {10.1086/513700},
archivePrefix = {arXiv},
       eprint = {astro-ph/0603449},
 primaryClass = {astro-ph},
       adsurl = {https://ui.adsabs.harvard.edu/abs/2007ApJS..170..377S}
}

@ARTICLE{2009ApJS..180..330K,
       author = {{Komatsu}, E. and {Dunkley}, J. and {Nolta}, M.~R. and {Bennett}, C.~L. and {Gold}, B. and {Hinshaw}, G. and {Jarosik}, N. and {Larson}, D. and {Limon}, M. and {Page}, L. and {Spergel}, D.~N. and {Halpern}, M. and {Hill}, R.~S. and {Kogut}, A. and {Meyer}, S.~S. and {Tucker}, G.~S. and {Weiland}, J.~L. and {Wollack}, E. and {Wright}, E.~L.},
        title = "{Five-Year Wilkinson Microwave Anisotropy Probe Observations: Cosmological Interpretation}",
      journal = {\apjs},
         year = 2009,
        month = feb,
       volume = {180},
       number = {2},
        pages = {330-376},
          doi = {10.1088/0067-0049/180/2/330},
archivePrefix = {arXiv},
       eprint = {0803.0547},
 primaryClass = {astro-ph},
       adsurl = {https://ui.adsabs.harvard.edu/abs/2009ApJS..180..330K}
}

@ARTICLE{2004PhRvD..69j3501T,
       author = {{Tegmark}, Max and {Strauss}, Michael A. and {Blanton}, Michael R. and {Abazajian}, Kevork and {Dodelson}, Scott and {Sandvik}, Havard and {Wang}, Xiaomin and {Weinberg}, David H. and {Zehavi}, Idit and {Bahcall}, Neta A. and {Hoyle}, Fiona and {Schlegel}, David and {Scoccimarro}, Roman and {Vogeley}, Michael S. and {Berlind}, Andreas and {Budavari}, Tam{\'a}s and {Connolly}, Andrew and {Eisenstein}, Daniel J. and {Finkbeiner}, Douglas and {Frieman}, Joshua A. and {Gunn}, James E. and {Hui}, Lam and {Jain}, Bhuvnesh and {Johnston}, David and {Kent}, Stephen and {Lin}, Huan and {Nakajima}, Reiko and {Nichol}, Robert C. and {Ostriker}, Jeremiah P. and {Pope}, Adrian and {Scranton}, Ryan and {Seljak}, Uro{\v{s}} and {Sheth}, Ravi K. and {Stebbins}, Albert and {Szalay}, Alexander S. and {Szapudi}, Istv{\'a}n and {Xu}, Yongzhong and {Annis}, James and {Brinkmann}, J. and {Burles}, Scott and {Castander}, Francisco J. and {Csabai}, Istvan and {Loveday}, Jon and {Doi}, Mamoru and {Fukugita}, Masataka and {Gillespie}, Bruce and {Hennessy}, Greg and {Hogg}, David W. and {Ivezi{\'c}}, {\v{Z}}eljko and {Knapp}, Gillian R. and {Lamb}, Don Q. and {Lee}, Brian C. and {Lupton}, Robert H. and {McKay}, Timothy A. and {Kunszt}, Peter and {Munn}, Jeffrey A. and {O'Connell}, Liam and {Peoples}, John and {Pier}, Jeffrey R. and {Richmond}, Michael and {Rockosi}, Constance and {Schneider}, Donald P. and {Stoughton}, Christopher and {Tucker}, Douglas L. and {vanden Berk}, Daniel E. and {Yanny}, Brian and {York}, Donald G.},
        title = "{Cosmological parameters from SDSS and WMAP}",
      journal = {\prd},
         year = 2004,
        month = may,
       volume = {69},
       number = {10},
          eid = {103501},
        pages = {103501},
          doi = {10.1103/PhysRevD.69.103501},
archivePrefix = {arXiv},
       eprint = {astro-ph/0310723},
 primaryClass = {astro-ph},
       adsurl = {https://ui.adsabs.harvard.edu/abs/2004PhRvD..69j3501T}
}

@ARTICLE{2005PhRvD..71j3515S,
       author = {{Seljak}, Uro{\v{s}} and {Makarov}, Alexey and {McDonald}, Patrick and {Anderson}, Scott F. and {Bahcall}, Neta A. and {Brinkmann}, J. and {Burles}, Scott and {Cen}, Renyue and {Doi}, Mamoru and {Gunn}, James E. and {Ivezi{\'c}}, {\v{Z}}eljko and {Kent}, Stephen and {Loveday}, Jon and {Lupton}, Robert H. and {Munn}, Jeffrey A. and {Nichol}, Robert C. and {Ostriker}, Jeremiah P. and {Schlegel}, David J. and {Schneider}, Donald P. and {Tegmark}, Max and {Berk}, Daniel E. and {Weinberg}, David H. and {York}, Donald G.},
        title = "{Cosmological parameter analysis including SDSS Ly{\ensuremath{\alpha}} forest and galaxy bias: Constraints on the primordial spectrum of fluctuations, neutrino mass, and dark energy}",
      journal = {\prd},
         year = 2005,
        month = may,
       volume = {71},
       number = {10},
          eid = {103515},
        pages = {103515},
          doi = {10.1103/PhysRevD.71.103515},
archivePrefix = {arXiv},
       eprint = {astro-ph/0407372},
 primaryClass = {astro-ph},
       adsurl = {https://ui.adsabs.harvard.edu/abs/2005PhRvD..71j3515S}
}

@ARTICLE{2019PhRvD.100j4027L,
       author = {{Lazkoz}, Ruth and {Lobo}, Francisco S.~N. and {Ortiz-Ba{\~n}os}, Mar{\'\i}a and {Salzano}, Vincenzo},
        title = "{Observational constraints of f (Q ) gravity}",
      journal = {\prd},
         year = 2019,
        month = nov,
       volume = {100},
       number = {10},
          eid = {104027},
        pages = {104027},
          doi = {10.1103/PhysRevD.100.104027},
archivePrefix = {arXiv},
       eprint = {1907.13219},
 primaryClass = {gr-qc},
       adsurl = {https://ui.adsabs.harvard.edu/abs/2019PhRvD.100j4027L}
}

@article{buchdahl1970non,
  title={Non-linear Lagrangians and cosmological theory},
  author={Buchdahl, Hans A},
  journal={Monthly Notices of the Royal Astronomical Society},
  volume={150},
  number={1},
  pages={1--8},
  year={1970},
  publisher={Oxford University Press Oxford, UK}
}

@article{harko2011f,
  title={f (R, T) gravity},
  author={Harko, Tiberiu and Lobo, Francisco SN and Nojiri, Shin’ichi and Odintsov, Sergei D},
  journal={Physical Review D—Particles, Fields, Gravitation, and Cosmology},
  volume={84},
  number={2},
  pages={024020},
  year={2011},
  publisher={APS}
}

@book{aldrovandi2012teleparallel,
    author = "Aldrovandi, Ruben and Pereira, Jos{\'e} Geraldo",
    title = "{Teleparallel Gravity}: {An Introduction}",
    doi = "10.1007/978-94-007-5143-9",
    isbn = "978-94-007-5142-2, 978-94-007-5143-9",
    publisher = "Springer",
    year = "2013"
}

@ARTICLE{2025ApJ...986..231R,
       author = {{Rubin}, David and {Aldering}, Greg and {Betoule}, Marc and {Fruchter}, Andy and {Huang}, Xiaosheng and {Kim}, Alex G. and {Lidman}, Chris and {Linder}, Eric and {Perlmutter}, Saul and {Ruiz-Lapuente}, Pilar and {Suzuki}, Nao},
        title = "{Union through UNITY: Cosmology with 2000 SNe Using a Unified Bayesian Framework}",
      journal = {\apj},
         year = 2025,
        month = jun,
       volume = {986},
       number = {2},
          eid = {231},
        pages = {231},
          doi = {10.3847/1538-4357/adc0a5},
archivePrefix = {arXiv},
       eprint = {2311.12098},
 primaryClass = {astro-ph.CO},
       adsurl = {https://ui.adsabs.harvard.edu/abs/2025ApJ...986..231R}
}

@ARTICLE{2020PhRvD.102l4029M,
       author = {{Mandal}, Sanjay and {Wang}, Deng and {Sahoo}, P.~K.},
        title = "{Cosmography in f (Q ) gravity}",
      journal = {\prd},
         year = 2020,
        month = dec,
       volume = {102},
       number = {12},
          eid = {124029},
        pages = {124029},
          doi = {10.1103/PhysRevD.102.124029},
archivePrefix = {arXiv},
       eprint = {2011.00420},
 primaryClass = {gr-qc},
       adsurl = {https://ui.adsabs.harvard.edu/abs/2020PhRvD.102l4029M}
}

@ARTICLE{2022EPJC...82..303Z,
       author = {{Zhao}, Dehao},
        title = "{Covariant formulation of f(Q) theory}",
      journal = {European Physical Journal C},
         year = 2022,
        month = apr,
       volume = {82},
       number = {4},
          eid = {303},
        pages = {303},
          doi = {10.1140/epjc/s10052-022-10266-4},
archivePrefix = {arXiv},
       eprint = {2104.02483},
 primaryClass = {gr-qc},
       adsurl = {https://ui.adsabs.harvard.edu/abs/2022EPJC...82..303Z}
}

@ARTICLE{2023MNRAS.522..252S,
       author = {{Sokoliuk}, Oleksii and {Arora}, Simran and {Praharaj}, Subhrat and {Baransky}, Alexander and {Sahoo}, P.~K.},
        title = "{On the impact of f(Q) gravity on the large scale structure}",
      journal = {\mnras},
         year = 2023,
        month = jun,
       volume = {522},
       number = {1},
        pages = {252-267},
          doi = {10.1093/mnras/stad968},
archivePrefix = {arXiv},
       eprint = {2303.17341},
 primaryClass = {astro-ph.CO},
       adsurl = {https://ui.adsabs.harvard.edu/abs/2023MNRAS.522..252S}
}

@ARTICLE{2019EPJC...79..530L,
       author = {{Lu}, Jianbo and {Zhao}, Xin and {Chee}, Guoying},
        title = "{Cosmology in symmetric teleparallel gravity and its dynamical system}",
      journal = {European Physical Journal C},
         year = 2019,
        month = jun,
       volume = {79},
       number = {6},
          eid = {530},
        pages = {530},
          doi = {10.1140/epjc/s10052-019-7038-3},
archivePrefix = {arXiv},
       eprint = {1906.08920},
 primaryClass = {gr-qc},
       adsurl = {https://ui.adsabs.harvard.edu/abs/2019EPJC...79..530L}
}

@ARTICLE{2021PhRvD.103f3505A,
       author = {{Ayuso}, Ismael and {Lazkoz}, Ruth and {Salzano}, Vincenzo},
        title = "{Observational constraints on cosmological solutions of f (Q ) theories}",
      journal = {\prd},
         year = 2021,
        month = mar,
       volume = {103},
       number = {6},
          eid = {063505},
        pages = {063505},
          doi = {10.1103/PhysRevD.103.063505},
archivePrefix = {arXiv},
       eprint = {2012.00046},
 primaryClass = {astro-ph.CO},
       adsurl = {https://ui.adsabs.harvard.edu/abs/2021PhRvD.103f3505A}
}

@ARTICLE{2022PhRvD.105h4061E,
       author = {{Esposito}, Fabrizio and {Carloni}, Sante and {Cianci}, Roberto and {Vignolo}, Stefano},
        title = "{Reconstructing isotropic and anisotropic f (Q ) cosmologies}",
      journal = {\prd},
         year = 2022,
        month = apr,
       volume = {105},
       number = {8},
          eid = {084061},
        pages = {084061},
          doi = {10.1103/PhysRevD.105.084061},
archivePrefix = {arXiv},
       eprint = {2107.14522},
 primaryClass = {gr-qc},
       adsurl = {https://ui.adsabs.harvard.edu/abs/2022PhRvD.105h4061E}
}

@ARTICLE{2023PhRvD.107d4022K,
       author = {{Khyllep}, Wompherdeiki and {Dutta}, Jibitesh and {Saridakis}, Emmanuel N. and {Yesmakhanova}, Kuralay},
        title = "{Cosmology in f (Q ) gravity: A unified dynamical systems analysis of the background and perturbations}",
      journal = {\prd},
         year = 2023,
        month = feb,
       volume = {107},
       number = {4},
          eid = {044022},
        pages = {044022},
          doi = {10.1103/PhysRevD.107.044022},
archivePrefix = {arXiv},
       eprint = {2207.02610},
 primaryClass = {gr-qc},
       adsurl = {https://ui.adsabs.harvard.edu/abs/2023PhRvD.107d4022K}
}

@ARTICLE{2021PhLB..82236634A,
       author = {{Anagnostopoulos}, Fotios K. and {Basilakos}, Spyros and {Saridakis}, Emmanuel N.},
        title = "{First evidence that non-metricity f(Q) gravity could challenge {\ensuremath{\Lambda}}CDM}",
      journal = {Physics Letters B},
         year = 2021,
        month = nov,
       volume = {822},
          eid = {136634},
        pages = {136634},
          doi = {10.1016/j.physletb.2021.136634},
archivePrefix = {arXiv},
       eprint = {2104.15123},
 primaryClass = {gr-qc},
       adsurl = {https://ui.adsabs.harvard.edu/abs/2021PhLB..82236634A}
}

@ARTICLE{2023EPJC...83...58A,
       author = {{Anagnostopoulos}, Fotios K. and {Gakis}, Viktor and {Saridakis}, Emmanuel N. and {Basilakos}, Spyros},
        title = "{New models and big bang nucleosynthesis constraints in f(Q) gravity}",
      journal = {European Physical Journal C},
         year = 2023,
        month = jan,
       volume = {83},
       number = {1},
          eid = {58},
        pages = {58},
          doi = {10.1140/epjc/s10052-023-11190-x},
archivePrefix = {arXiv},
       eprint = {2205.11445},
 primaryClass = {gr-qc},
       adsurl = {https://ui.adsabs.harvard.edu/abs/2023EPJC...83...58A}
}

@ARTICLE{2017ApJ...838L..15L,
       author = {{Lin}, H. and {Buckley-Geer}, E. and {Agnello}, A. and {Ostrovski}, F. and {McMahon}, R.~G. and {Nord}, B. and {Kuropatkin}, N. and {Tucker}, D.~L. and {Treu}, T. and {Chan}, J.~H.~H. and {Suyu}, S.~H. and {Diehl}, H.~T. and {Collett}, T. and {Gill}, M.~S.~S. and {More}, A. and {Amara}, A. and {Auger}, M.~W. and {Courbin}, F. and {Fassnacht}, C.~D. and {Frieman}, J. and {Marshall}, P.~J. and {Meylan}, G. and {Rusu}, C.~E. and {Abbott}, T.~M.~C. and {Abdalla}, F.~B. and {Allam}, S. and {Banerji}, M. and {Bechtol}, K. and {Benoit-L{\'e}vy}, A. and {Bertin}, E. and {Brooks}, D. and {Burke}, D.~L. and {Carnero Rosell}, A. and {Carrasco Kind}, M. and {Carretero}, J. and {Castander}, F.~J. and {Crocce}, M. and {D'Andrea}, C.~B. and {da Costa}, L.~N. and {Desai}, S. and {Dietrich}, J.~P. and {Eifler}, T.~F. and {Finley}, D.~A. and {Flaugher}, B. and {Fosalba}, P. and {Garc{\'\i}a-Bellido}, J. and {Gaztanaga}, E. and {Gerdes}, D.~W. and {Goldstein}, D.~A. and {Gruen}, D. and {Gruendl}, R.~A. and {Gschwend}, J. and {Gutierrez}, G. and {Honscheid}, K. and {James}, D.~J. and {Kuehn}, K. and {Lahav}, O. and {Li}, T.~S. and {Lima}, M. and {Maia}, M.~A.~G. and {March}, M. and {Marshall}, J.~L. and {Martini}, P. and {Melchior}, P. and {Menanteau}, F. and {Miquel}, R. and {Ogando}, R.~L.~C. and {Plazas}, A.~A. and {Romer}, A.~K. and {Sanchez}, E. and {Schindler}, R. and {Schubnell}, M. and {Sevilla-Noarbe}, I. and {Smith}, M. and {Smith}, R.~C. and {Sobreira}, F. and {Suchyta}, E. and {Swanson}, M.~E.~C. and {Tarle}, G. and {Thomas}, D. and {Walker}, A.~R. and {DES Collaboration}},
        title = "{Discovery of the Lensed Quasar System DES J0408-5354}",
      journal = {\apjl},
         year = 2017,
        month = apr,
       volume = {838},
       number = {2},
          eid = {L15},
        pages = {L15},
          doi = {10.3847/2041-8213/aa624e},
archivePrefix = {arXiv},
       eprint = {1702.00072},
 primaryClass = {astro-ph.GA},
       adsurl = {https://ui.adsabs.harvard.edu/abs/2017ApJ...838L..15L}
}

@ARTICLE{2020MNRAS.494.6072S,
       author = {{Shajib}, A.~J. and {Birrer}, S. and {Treu}, T. and {Agnello}, A. and {Buckley-Geer}, E.~J. and {Chan}, J.~H.~H. and {Christensen}, L. and {Lemon}, C. and {Lin}, H. and {Millon}, M. and {Poh}, J. and {Rusu}, C.~E. and {Sluse}, D. and {Spiniello}, C. and {Chen}, G.~C.-F. and {Collett}, T. and {Courbin}, F. and {Fassnacht}, C.~D. and {Frieman}, J. and {Galan}, A. and {Gilman}, D. and {More}, A. and {Anguita}, T. and {Auger}, M.~W. and {Bonvin}, V. and {McMahon}, R. and {Meylan}, G. and {Wong}, K.~C. and {Abbott}, T.~M.~C. and {Annis}, J. and {Avila}, S. and {Bechtol}, K. and {Brooks}, D. and {Brout}, D. and {Burke}, D.~L. and {Carnero Rosell}, A. and {Carrasco Kind}, M. and {Carretero}, J. and {Castander}, F.~J. and {Costanzi}, M. and {da Costa}, L.~N. and {De Vicente}, J. and {Desai}, S. and {Dietrich}, J.~P. and {Doel}, P. and {Drlica-Wagner}, A. and {Evrard}, A.~E. and {Finley}, D.~A. and {Flaugher}, B. and {Fosalba}, P. and {Garc{\'\i}a-Bellido}, J. and {Gerdes}, D.~W. and {Gruen}, D. and {Gruendl}, R.~A. and {Gschwend}, J. and {Gutierrez}, G. and {Hollowood}, D.~L. and {Honscheid}, K. and {Huterer}, D. and {James}, D.~J. and {Jeltema}, T. and {Krause}, E. and {Kuropatkin}, N. and {Li}, T.~S. and {Lima}, M. and {MacCrann}, N. and {Maia}, M.~A.~G. and {Marshall}, J.~L. and {Melchior}, P. and {Miquel}, R. and {Ogando}, R.~L.~C. and {Palmese}, A. and {Paz-Chinch{\'o}n}, F. and {Plazas}, A.~A. and {Romer}, A.~K. and {Roodman}, A. and {Sako}, M. and {Sanchez}, E. and {Santiago}, B. and {Scarpine}, V. and {Schubnell}, M. and {Scolnic}, D. and {Serrano}, S. and {Sevilla-Noarbe}, I. and {Smith}, M. and {Soares-Santos}, M. and {Suchyta}, E. and {Tarle}, G. and {Thomas}, D. and {Walker}, A.~R. and {Zhang}, Y.},
        title = "{STRIDES: a 3.9 per cent measurement of the Hubble constant from the strong lens system DES J0408-5354}",
      journal = {\mnras},
         year = 2020,
        month = jun,
       volume = {494},
       number = {4},
        pages = {6072-6102},
          doi = {10.1093/mnras/staa828},
archivePrefix = {arXiv},
       eprint = {1910.06306},
 primaryClass = {astro-ph.CO},
       adsurl = {https://ui.adsabs.harvard.edu/abs/2020MNRAS.494.6072S}
}

@article{bhardwaj2024cosmological,
  title={Cosmological dynamics of accelerating model in f (Q) gravity with latest observational data},
  author={Bhardwaj, Vinod Kumar and Garg, Priyanka and Prakash, Suraj},
  journal={Astrophysics and Space Science},
  volume={369},
  number={5},
  pages={50},
  year={2024},
  publisher={Springer}
}
